\documentclass[iop]{emulateapj}

\usepackage{natbib}
\usepackage{graphicx}
\usepackage{subfigure}
\usepackage{amsmath}
\usepackage[utf8]{inputenc}
\usepackage{float}
\usepackage [english]{babel}
\usepackage [autostyle, english = american]{csquotes}
\MakeOuterQuote{"}

\usepackage{multirow}
\usepackage{array}
\newcolumntype{M}[1]{>{\centering\arraybackslash}m{#1}}
\newcolumntype{N}{@{}m{0pt}@{}}

\shorttitle{MMR, Exoplanet Detection \& Characterization}
\shortauthors{Tabeshian \& Wiegert}

\def\las{\mathrel{\hbox{\rlap{\hbox{\lower3pt\hbox{$\sim$}}}\hbox{\raise2pt\hbox{$<$}}}}}
\def\gas{\mathrel{\hbox{\rlap{\hbox{\lower3pt\hbox{$\sim$}}}\hbox{\raise2pt\hbox{$>$}}}}}

\begin{document}

\title{Detection and Characterization of Extrasolar Planets through Mean-Motion Resonances. II. The Effect of the Planet's Orbital Eccentricity on Debris Disk Structures}

\author{Maryam Tabeshian\altaffilmark{1} and Paul A. Wiegert\altaffilmark{1,2}}
\affil{$^1$Department of Physics and Astronomy, The University of Western Ontario, London, ON, Canada, N6A 3K7 \\ 
$^2$Centre for Planetary Science and Exploration, The University of Western Ontario, London, ON, Canada, N6A 3K7 \\ \\
\textit{Published in The Astrophysical Journal (ApJ, 847, 24), September 15, 2017}}

\email{mtabeshi@uwo.ca}

\begin{abstract}
Structures observed in debris disks may be caused by gravitational interaction with planetary or stellar companions. These perturbed disks are often thought to indicate the presence of planets and offer insights into the properties of both the disk and the perturbing planets. Gaps in debris disks may indicate a planet physically present within the gap, but such gaps can also occur away from the planet's orbit at mean-motion resonances (MMRs), and this is the focus of our interest here. We extend our study of planet–disk interaction through MMRs, presented in an earlier paper, to systems in which the perturbing planet has moderate orbital eccentricity, a common occurrence in exoplanetary systems. In particular, a new result is that the 3:1 MMR becomes distinct at higher eccentricity, while its effects are absent for circular planetary orbits. We also only consider gravitational interaction with a planetary body of at least $1 M_J$. Our earlier work shows that even a 1 Earth mass planet can theoretically open an MMR gap; however, given the narrow gap that can be opened by a low-mass planet, its observability would be questionable. We find that the widths, locations, and shapes of two prominent structures, the 2:1 and 3:1 MMRs, could be used to determine the mass, semimajor axis, and eccentricity of the planetary perturber and present an algorithm for doing so. These MMR structures can be used to narrow the position and even determine the planetary properties (such as mass) of any inferred but as-yet-unseen planets within a debris disk. We also briefly discuss the implications of eccentric disks on brightness asymmetries and their dependence on the wavelengths with which these disks are observed. \\
\end{abstract}



\section{Introduction}
\label{Sec:Intro}

Stars that exhibit excess infrared emission are thought to host disks of circumstellar material known as "debris disks," with the extra emission being linked to heating and reradiation by the constituent particles. Such disks are believed to be remnants of the cloud of gas and dust that formed the star but that have not coagulated to form planets. Particles that make up debris disks range in size from planetesimals with radii of a few hundreds of kilometers to small submicron-size dust grains. The infrared excess comes from the smaller grains, which produce a second "bump" in the star's blackbody curve. In this manner, in 1984, a debris disk was discovered outside the solar system for the first time. Using data from \textit{the Infrared Astronomical Satellite (IRAS)}, the spectral energy distribution of the main-sequence star Vega was found to have an infrared excess. Detailed analysis indicated the presence of circumstellar dust particles with radii greater than a millimeter and a mean distance of 85 AU from the star \citep{Aumann84}. Since then, many extrasolar debris disks have been detected using optical to submillimeter observations \citep[see the review paper by][]{Wyatt08}. 

Particles in the smallest end of the size distribution (submicron) are blown away by stellar radiation pressure over very short (orbital) timescales. At slightly larger sizes, the Poynting–Robertson (PR) drag is effective in removing dust grains on bound orbits with $1~ \mu m < s < 1~ mm$ over timescales of $\sim10^4$ years \citep{Klacka08}. However, this effect can safely be ignored when studying debris disks, since the timescale for the inspiraling of dust due to the PR drag is typically longer than its collisional lifetime and therefore particles are collisionally ground down to small sizes and blown out of the system by stellar radiation before spiraling inward \citep{Wyatt99}. The justifications for neglecting radiation effects in our simulations will be examined in more detail in Section \ref{Sec:PRvscoll}.

A planet in the vicinity of a debris disk may leave its signature in the disk structure. Before the gas in a protoplanetary disk is blown away by stellar radiation, drag against the gas causes dust particles to settle into circular orbits in the same plane as the plane of rotation of the star. However, if the system contains a planet on an elliptical orbit, particle orbits are perturbed such that the disk loses its circular shape and the center of its symmetry becomes offset from the star (see Section \ref{Sec:PeriAlign_Offset}). Moreover, if the planet is not in the same orbital plane as the disk, its dynamical effect on the disk particles may reveal itself as a warp in the disk. For instance, the warped inner disk around the star $\beta$ Pictoris, first noted by \cite{Burrows95} through imaging in optical scattered light using the \textit{Hubble Space Telescope} and followed up by ground-based adaptive optics observations in the near-infrared by \cite{Mouillet97}, was believed to be an indication of dynamical interaction with a previously undetected planet on an inclined orbit to the disk particles \citep[see][]{Augereau01}. This hypothesis was later endorsed and confirmed when direct imaging revealed an inclined planet just outside the innermost belt of $\beta$ Pic \citep{Lagrange10}. 

In addition to offsets and warps, which are believed to be caused by the perturber's orbital eccentricity and inclination, structures can be formed in the disk through the particles' gravitational interactions with the planet via mean-motion resonances (MMRs). Two bodies are said to be in MMR if their orbits are commensurable, meaning that for every $p$ number of times one of them orbits the star, the other completes $p+q$ orbits, where $p$ and $q$ are positive integers, with the latter ($q$) defining the order of the resonance. This could have one of two consequences for the orbit of the less massive body: if it avoids a close encounter with the planet at conjunction, a stable resonance is formed, otherwise the repeated perturbation of the orbit of the less massive body results in a change of its orbital elements.

If the former scenario occurs, particles accumulate at MMRs. In fact, dust density enhancements have been observed in debris disks and are attributed to the trapping of dust particles in exterior MMRs with a planet. This was suggested theoretically by \cite{Gold75}, who proposed that as interplanetary dust spirals inward due to the PR drag, dust particles can fall into MMR with a planet interior to their orbits. This could make their orbits temporarily stable despite the PR effect, and their lifetimes can be extended by a factor of a few up to $\sim$ 100,000 yrs \citep{Jackson89}. In this case \cite{Gold75} argued that ringlike circumstellar structures with particle densities on the order of $10^4$ times larger than average could be formed. This also explains the stability of some dust particles in the solar system's zodiacal cloud, observed in the \textit{IRAS} and \textit{Cosmic Background Explorer (COBE)} data, where particles are trapped by the Earth \citep{Dermott94}.

Alternatively, \cite{Wyatt03} suggested that planet migration at an earlier stage in the life of a planetary system could also trap particles in resonances and result in the formation of clumpy structures in debris disks. \cite{Wyatt03} proposed that the migration history of a system can be understood by studying the planet's signature in the observed spatial distribution of particles in the debris disk. This scenario has been used to explain the capture of Pluto and the Plutinos in Neptune's 3:2 MMR. The outward migration of Neptune's orbit could have resulted in the trapping of smaller bodies in exterior 3:2 MMR \citep[see, for instance,][]{Malhotra95, Hahn99}.

Density enhancements due to resonant dust trapping might have been observed in some extrasolar dust disks. For instance, submillimeter observations of the debris disk around $\epsilon$ Eridani using the James Clerk Maxwell Telescope showed a dusty ring 60 AU from the star with four emission peaks \citep{Greaves98}. These could be explained as structures formed by dust particles captured in 3:2 exterior MMR with an $\sim$0.2 Jupiter-mass ($M_J$) planet on a circular orbit 55–65 AU from the star \citep{Ozernoy00}. Later works have come to somewhat different conclusions (the mass, semimajor axis, and eccentricity of the planet have been determined to be 0.1 $M_J$, 40 AU and 0.3, respectively, while the four peaks of emission have been attributed to the trapping of dust particles in 3:2 and 5:3 exterior MMRs with the planet at periastron \citep{Quillen02}); nevertheless, the existence of resonant structures in disks has not been disputed. In another example, \cite{Wilner02} detected two dust emission peaks in Vega's dusty disk using \textit{the Institute for Radio Astronomy in the Millimeter Range (IRAM)} interferometer at 1.3 mm wavelength. They attributed this to the trapping of dust, via dust migration under PR drag, into 2:1 and 3:1 resonances with a 3 $M_J$ planet having an orbital eccentricity $e=0.6$. Again in this case, alternative resonant models were shown to be consistent with observations. \cite{Wyatt03} proposed that the observed dust overdensity is due to particles trapped in the 2:1 and 3:2 MMRs by migration of a Neptune-mass planet from 40 to 65 AU over a period of 56 Myr. Regardless of the capture mechanism, the formation of dust density structures due to MMRs can be taken as an indication of the presence of a planet whose mass and orbital parameters may be determinable through the properties of the affected dust population.

Whereas resonant dust trapping and its implication for planet detection and characterization has been discussed to some extent in the literature, not much emphasis has been placed on understanding structures formed by resonant gap formation in extrasolar debris disks, which is the focus of the present study in an attempt to understand how such resonant gaps could be diagnostic of planetary parameters without the need to observe the planet itself.

The location of each resonance can be found analytically by simply considering the definition of MMR, which occurs when the mean motion of one particle is a simple fraction of that of the other. Expressed in terms of the two particles' semimajor axes, $a$ and $a ^\prime$, the resonance location for any $p$ and $q$ combination is expressed by Equation \ref{Eq:MMR_Def_a},

\begin{equation}
\label{Eq:MMR_Def_a}
a^\prime = \left(\frac{p+q}{p}\right)^{\frac{2}{3}k} ~ a ~,
\end{equation}

\noindent where $k=+1$ for exterior resonance (i.e., $a^\prime > a$) and $k=-1$ for interior resonance (i.e., $a^\prime < a$). Note that we adopt the same notation we used in our earlier paper (\cite{Tabeshian16}, hereafter \textit{Paper I}), where the primed and unprimed quantities denote the orbital elements of the particle being perturbed (the "asteroid") and the perturbing body ("the planet"), respectively. Also note that here we are interested in the effect of MMRs at locations specifically away from the planet's orbit; thus, we do not investigate the 1:1 MMR, nor are we concerned with the gap clearing that occurs in the feeding zone of a planet due to its tidal interaction with the disk.

Periodic perturbations of a particle's orbit by a more massive one can eventually result in significant changes in the orbit of the less massive body. Such perturbations cause the resonant argument of the disturbing function, $\phi$ defined by Equation \ref{Eq:phi}, to librate about a fixed value. If the amplitude of the libration becomes large enough, either the MMRs could eventually remove the less massive particle from its orbit or the particle could stay in a bound orbit but could gain some eccentricity. In either case, a gap develops due to the particles having been ejected from their orbits or having developed a large radial excursion that makes them spend a large fraction of their orbital period away from the location of the resonance. The resonant angle $\phi$ is given by \cite{Murray99} as

\begin{equation}
\label{Eq:phi}
\phi = j_1 \lambda^\prime + j_2 \lambda + j_3 \varpi^\prime + j_4 \varpi ~,
\end{equation}

\noindent where $j_1=p+q$, $j_2=-p$, and $j_3$ and $j_4$ are either zero or $-q$ depending on the relative locations of the perturbing body and the one being perturbed, while $\lambda$ and $\lambda^\prime$ are the mean longitudes and $\varpi$ and $\varpi^\prime$ are the longitudes of pericenter. Therefore, the resonant argument, $\phi$, defines the angle between the longitude of the conjunction of the two bodies and the longitude of the pericenter of the object with the larger semimajor axis. To lowest order in eccentricity, the maximum libration width, $\delta a^{\prime}_{max}$ at each first- and second-order resonance can also be calculated using Equations (8.76) and (8.58) of \cite{Murray99}.

In debris disks, resonant interaction with planets could cause the formation of gaps that may or may not be observable telescopically depending on disk particle eccentricities, as will be discussed briefly in Section \ref{Sec:Results}. Such gaps have been observed in numerical simulations of debris disks by some authors but have not been extensively studied. For instance, in an attempt to understand gravitational sculpting of a single planet orbiting interior to the Fomalhaut disk, \cite{Chiang09} showed that resonance gaps could form for a variety of planet mass–semimajor axis combinations. However, the authors do not take the discussion further to describe how such structures would help characterize the planet causing them. Furthermore, simulations by \cite{Nesvold15} show a gap in the disk's surface brightness distribution at the 2:1 MMR with a 3 $M_J$ planet at 50 AU. Though the authors address the depletion of planetesimals at this resonance, they do not discuss how it can yield measurements of planetary parameters. Similarly, in \cite{Reche08}, an example of a nonmigrating planet interior to a simulated disk that has three gaps whose locations correspond to the 3:2 and 2:1 exterior MMRs with the planet is shown, but they are not addressed by the authors. Nevertheless, the location and appearance of the 2:1 gap in their figure resemble our results in this paper. These illustrate the rich variety of structures that can be created by resonances and the need to understand this process and what it tells us about the system.

This work focuses on structures formed by resonant gap formation in debris disks through gravitational interaction with a single nonmigrating planet and the consequent formation of what would be analogous to the solar system's Kirkwood gaps \citep{Kirkwood67}. Our solar system's Kirkwood gaps are complicated by multiplanet effects such as secular resonances, but we leave to future work the study of MMR gap formation in planetary systems with more than one planet.

We argue that under certain conditions, the gaps discussed in this paper could be visible in telescopic images of debris disks. We showed in \textit{Paper I} that dynamical interactions of a single planet with a gas-poor and dynamically cold planetesimal disk can result in the formation of azimuthally asymmetric gaps whose widths and locations are diagnostic of the perturber's mass and semimajor axis, even if the planet remains unseen. We restricted our analysis to systems in which the perturbing planet was either on a circular orbit or had a small, ~0.05 orbital eccentricity. However, unlike planets in our own solar system, most exoplanets found to date have significant eccentricities \citep[see the review paper by][]{Winn15}. Therefore, here we extend our analysis of planet–disk interactions and resonance gap formation to systems with the planet on a range of higher-eccentricity orbits in order to provide a more complete picture of the gap formation that results from resonant interactions between a single planet and a planetesimal disk. Here we study the dynamic structures of planetesimals in dynamically cold systems and assume that radiation and PR drag forces can be neglected, as will be addressed in more detail in Section \ref{Sec:PRvscoll} and following a similar treatment in our previous work. We find that MMR gap structures would be detectable in telescopic images of disk systems hosting planets on moderately eccentric orbits, though the resulting disk structures are more complex than in the low-eccentricity case.

We start this paper by describing in Section \ref{Sec:Dynamics} how the disk is dynamically affected by a planet on a noncircular orbit, and we also discuss the importance of radiation forces in debris disks. For our simulations, we use the same numerical method as in \textit{Paper I} (which we go over briefly in Section \ref{Sec:Sims}), but our initial conditions are different, appropriate to a debris disk with a planet on an eccentric orbit. We present our results for both interior and exterior MMRs in Section \ref{Sec:Results} and discuss their implications in Section \ref{Sec:Discussions}, where we also show how our simulated disks would look if observed by the \textit{Atacama Large Millimeter/submillimeter Array (ALMA)}. We end by a summary and conclusions in Section \ref{Sec:SummConc}.


\section{Disk Dynamics}
\label{Sec:Dynamics}

Here we will assume for simplicity a quiescent and dynamically cold disk perturbed by a single nonmigrating planet. However, the addition of a massive eccentric planet means that the particles in the disk cannot travel on perfectly circular orbits but are forced to take on minimally eccentric orbits. Such a disk, which corresponds to particles with a forced eccentricity but no free eccentricity, will be briefly outlined.


\subsection{Forced Eccentricity and Longitude of Pericenter}
\label{Sec:Forced_e}

When the perturbing planet has nonzero orbital eccentricity, it imposes an eccentricity onto the disk particles. This is referred to as the particles' "forced eccentricity" ($e_f^\prime$) and under the secular approximation is given by \citep{Murray99}

\begin{equation}
\label{Eq:Forced_e}
e_f^\prime (a^\prime) = \frac{b_{3/2}^{(2)} (\alpha)}{b_{3/2}^{(1)} (\alpha)} e ~,
\end{equation}

\noindent where $\alpha = (\frac{a}{a^\prime})^{k}$ while the $b_s^j (\alpha)$ terms are the Laplace coefficients given by \citep{Murray99}

\begin{equation}
\label{Eq:Laplace_Coeff}
b_s^j (\alpha) = \frac{1}{\pi} \int_{0}^{2\pi} \frac{\cos(j\psi)}{(1 - 2 ~ \alpha ~ \cos(\psi)+\alpha^2)^s} d\psi ~.
\end{equation}

Equation \ref{Eq:Forced_e} is independent of the perturber's mass and applies if the system contains only one perturber. The forced eccentricity in such systems is also independent of time, as the particles do not undergo any secular evolution \citep{Wyatt99}. Therefore, forced eccentricity only depends on the eccentricity of the perturber and diminishes with distance from it. The particles could in principle have an additional component of eccentricity, called the "free eccentricity," but we will take this to be zero, as is appropriate to a dynamically cold disk. In this case, the line of apses of the particles, $\varpi ^\prime$, aligns parallel to that of the planet.

In this study, we determine the maximum libration width at each MMR location by considering the forced eccentricity induced by a planet at the semimajor axis of each resonance, assuming that disk particles have negligible free eccentricity. Also, the disk particles are taken to orbit the star in the same plane as the planet; therefore, we do not consider warps in the disk that could be caused by the forced inclination of the planet.


\subsection{Disk Offset}
\label{Sec:PeriAlign_Offset}

The result of the eccentricity of the particles together with the alignment of their orbits with that of the planet is that the disk's center of symmetry is offset from the star by an amount that is related to the forced eccentricity of the particles' orbits. This causes an azimuthal brightness asymmetry in the disk.

The observed brightness asymmetry in the circumstellar disk around HR 4796A was first postulated to be due to an offset caused by gravitational perturbations of the disk by a low-mass stellar companion or a planet on a noncircular orbit \citep{Telesco00}. The nearly edge-on disk was found to be $5.9 \% \pm 3.2 \%$ brighter in the northeast lobe than it was in the southwest lobe, with a statistical significance of $1.8 \sigma$. Dynamical modeling of the HR4796A disk by \cite{Wyatt99} suggested that a stellar or planetary companion that could impose a forced eccentricity as small as 0.02 on the disk could cause the disk's center of symmetry to be offset from the star by $\sim$ 2 AU and could, therefore, be responsible for its observed brightness asymmetry. According to their model, in the absence of a stellar companion, a single planet with a mass of $> 10 ~M_\oplus$ and an eccentricity of $> 0.02$ could result in a $5 \%$ brightness asymmetry in the HR4796A disk, indicating that planets of moderate eccentricity could cause measurable offsets in debris disks. This phenomenon was dubbed the "pericenter glow" (since it results in the side of the disk closest to the star becoming warmer, and hence brighter, than the other) and has since been observed in other debris disks, such as the dusty disk around Fomalhaut, a disk that has an azimuthal brightness asymmetry and is offset by 15.3 AU \citep{Kalas05}. The offset in the center of symmetry of debris disks is now known to be a signature of a companion, stellar or planetary, that is on an eccentric orbit and forces the orbital elements of disk particles into pericenter alignment. The resulting pericenter glow is, therefore, caused by the disk being closer to the star on the forced pericenter side and hence warmer and brighter in the wavelength of observation.

In a recent study, \cite{Pan16} showed that azimuthal temperature asymmetries due to disk offset could be compensated by azimuthal asymmetries in dust density. At apocenter, particles travel more slowly and hence spend more time there, and so higher particle densities at apocenter could result in an apocenter glow instead. The authors argued that the apocenter/pericenter flux ratio is dependent on the wavelength of observation and suggested, through numerical modeling of the debris disks around Fomalhaut and $\epsilon$ Eridani, that apocenter glow wins over the enhanced flux due to disk offset away from pericenter if observed at far-infrared and submillimeter wavelengths, while the opposite happens at shorter wavelengths. We will revisit disk offset and apocenter/pericenter glow in Section \ref{Sec:DiskOffset}.


\subsection{The Importance of Radiation Forces in Debris Disks}
\label{Sec:PRvscoll}

In addition to the gravity of the star and the planet, important dynamical effects may in principle arise in debris disks from stellar radiation pressure and PR drag. Here we will ignore these radiative effects, partly for simplicity and partly because there is a wide range of debris disks for which these effects will be negligible. This point has been argued more thoroughly elsewhere \citep[see][]{Wyatt99}, but it is important enough to our study that the main points will be reviewed here as an order-of-magnitude calculation.

Radiative effects can be parameterized by the ratio of the radiation force to the star's gravitational force ($\beta$), which is a constant for a particular particle of a particular radius $r_d$. Extending the results of \cite{Weidenschilling93} to stars of different luminosities ($L_*$), the two are related through $\beta = \frac{0.57 \times 10^{-6}}{r_d}\frac{L_*}{L_\odot}$ at a density of 1000 kg/m$^{3}$, where $r_d$ is in meters.

There are three broad classes of particle behaviors based on their size. The smallest particles ($r_d \las 0.57~\mu$m) are blown out of the system by radiation pressure. Intermediate-sized particles spiral into the star under PR drag, while the largest particles ($r_d \gtrsim 500~\mu m$) are essentially unaffected. For example, a dust particle of radius $500~\mu m$ has a PR inspiral timescale from 100~AU around a Sun-like star of order a Hubble time. The effects of radiation forces on such large particles can safely be ignored; it is in the intermediate particles where they are most pronounced.

In a debris disk with a power-law size distribution like that of a collisional cascade, $r^{-3.5}$ \citep{Dohnanyi69}, most of the mass is in the largest objects (i.e., asteroids or planetesimals). However, the smaller dust particles will dominate the emission and hence the observations. We expect the emission to be dominated by particles of sizes comparable to the wavelength at which we are observing. As a result, the above determination that PR drag is most effective on particles sizes from roughly 1 to 500 $\mu m$ means that radiation forces cannot be dismissed out of hand.

The first effect of radiation forces on particles in this size range ($0.1 < \beta < 0.5$) is that, even if released from parents on nearly circular orbits, they are immediately placed on higher-eccentricity paths \citep{Kuchner10, Thebault12}. This can smear out spatial structures in the disk. For a Sun-like star, this $\beta$ range corresponds to sizes of 1 to several $\mu m$, and thus collisionally produced dust will be a complicating factor for observations through the mid-infrared. Disk structures should still be observable in longer wavelengths if the second effect of radiation forces, namely PR drag, is small enough. We turn our attention to this phenomenon next.

PR drag causes particles to spiral into the star but will have a negligible effect on our simulations if the collisional lifetime of dust is short compared to the PR timescale. That is, if interparticle collisions reduce the dust to small particles (which will be blown out of the system by radiation pressure on a very short timescale, essentially the orbital timescale) quickly enough, they will not drift far enough to smear out any disk structures. Here we argue that the collisional lifetime of dust will be much shorter than the PR timescale in many (though not all) physically realistic debris disks.

For the PR timescale, \cite{Weidenschilling93} calculated that for particles on near-circular orbits, the rate of change of the heliocentric radius, $R$, is

\begin{equation}
\frac{dR}{dt} = \frac{G M_* \beta}{R c} ~,
\end{equation} 

\noindent where $G$ is the Universal Gravitational Constant, $M_*$ is the star's mass, and $c$ is the speed of light. From this, the time to spiral into the star under the PR effect is just

\begin{eqnarray}
t_{PR} &=& \frac{R^2 c}{G M_* \beta} ~.
\end{eqnarray}

The collision timescale is more complicated. Here we will assume a power-law size distribution of the disk material like that of a collisional cascade \citep[see][]{Dohnanyi69}, with most of the mass in the large bodies (the asteroids) that are unaffected by the PR drag. The dust observed telescopically from Earth is continuously regenerated by asteroid collisions. A dust particle of radius $r_d$ cannot be disrupted by a collision with a particle much smaller than itself and collides only rarely with particles larger than itself. Thus, it is most likely collisionally disrupted in a collision with a particle of roughly its own size or somewhat greater \citep[see, for instance,][]{Wyatt99}. We approximate its collisional lifetime as the time it takes to sweep out a volume that should include one other particle its own size,

\begin{equation}
t_{coll} = \frac{1}{n_d \pi r_d^2 v_{rel}} ~,
\end{equation}

\noindent where $v_{rel}$ is the particle's relative velocity, that is, the velocity above that of a purely circular orbit $v_{orb} = \sqrt{\frac{G M_*}{R}}$. The number density of dust particles, $n_d$, is the dust production rate, $q_d$, times the length of time dust survives, all divided by the disk volume $V \sim \pi \frac{v_{rel}}{v_{orb}} R_d^3$. If we consider survival against collisions only, the survival time is $t_{coll}$, and $n_d = q_d t_{coll}/V$.

The dust production rate $q_d$ is $Q$ times the rate of asteroid collisions, $c_a$, times the relative asteroid and dust masses, where $Q$ is the fraction of asteroid mass converted to dust per asteroid collision. Let us assume all asteroids are the same size, $r_a$, for simplicity. Then the collision timescale for asteroids will be

\begin{equation}
t_{coll,a} = \frac{1}{n_a \pi r_a^2 v_{rel}} ~,
\end{equation}

\noindent where $n_a$ is the number density of asteroids, $N_a/V$. Here $N_a = \frac{3 M_{d}}{4 \pi r_a^3 \rho}$ is the total number of asteroids in a disk with mass $M_{d}$. From this, the dust production rate is

\begin{equation}
q_d = Q \left( \frac{r_a}{r_d} \right)^3 c_a = Q N_a n_a \pi r_a^2 v_{rel} \left( \frac{r_a}{r_d} \right)^3 ~,
\end{equation}

\noindent and the dust number density is

\begin{equation}
n_d =  Q N_a n_a \pi r_a^2 v_{rel} t_{coll} \left( \frac{r_a}{r_d} \right)^3 /V ~.
\end{equation}

Putting this back into our expression for $t_{coll}$, making the needed substitutions and rearranging, we get

\begin{equation}
t_{coll} \sim  \frac{4 \pi R_d^3\rho}{3 M_d} \sqrt{\frac{r_a r_d}{Q G M_*}}R^{0.5} ~.
\end{equation}

\begin{figure*}
    \centering
    \includegraphics[totalheight=0.4\textheight]{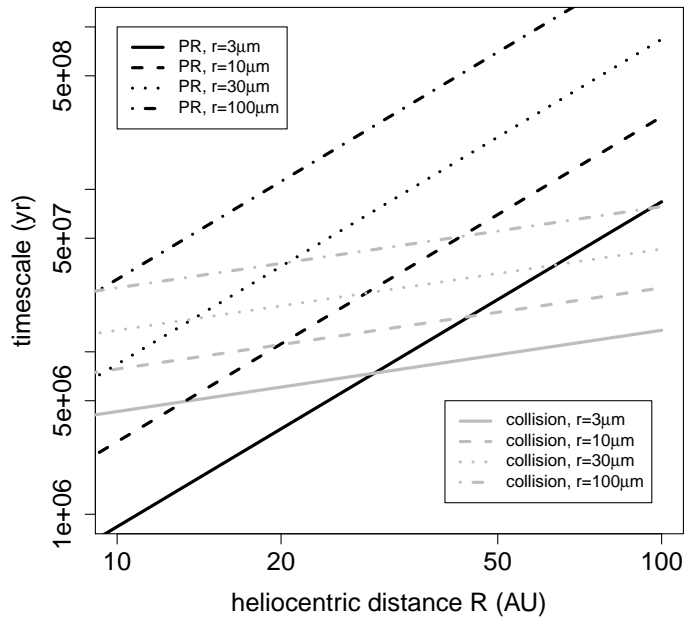}
    \caption{Comparison of the PR inspiral timescales to the collisional lifetimes of dust particles in a debris disk. See the text for details.}
\label{Fig:PRvscoll}
\end{figure*}

In Figure \ref{Fig:PRvscoll}, we plot the results for an asteroid of size $r_a$ = 1~km, a star with mass $M_* = 1 ~M_\odot$, an asteroid and dust density $\rho = 1000$~kg~m$^{-3}$, a dust production fraction $Q= 0.01$, and a disk with mass $M_{d} = 1$~Earth-mass and radius $R_d = 100$~AU. Four different particle sizes are considered from $r_d =$ 3 to 100~$\mu$m. The collisional lifetime of the dust is much shorter than the PR lifetime over most of the disk, though they cross over in the inner regions. Thus, we expect that PR drag will be less important in the outer regions of debris disks and when observations are taken at longer wavelengths. It is on the basis of this result that we choose to neglect radiation forces in this first look at the dynamical effects of planets on debris disks.

We do not claim that the PR drag is unimportant in all regions of all debris disks; indeed, it is likely to be important in some physically realistic disks, particularly those of low mass where the dust component is sparse (and hence the collisional lifetime of dust is very long). However, such dust-poor disks are also likely to be fainter. Here we choose to study the brighter and simpler disks; we recognize the possible importance of the PR drag in some cases, particularly when observing at shorter wavelengths, but leave that work to another study.

\section{Simulations}
\label{Sec:Sims}

\subsection{The Method}
\label{Sec:Method}

The symplectic integrator used for our simulations is the same as that in \textit{Paper I} and is based on the Wisdom–Holman algorithm \citep{Wisdom91}. We examine two possible cases: either interior or exterior MMRs in a disk of planetesimals that interact gravitationally with a single planet orbiting a 1 $M_\odot$ star. For simplicity, we ignore gravitational interaction among disk particles, as well as the radiation pressure and PR drag. Furthermore, we assume that little or no gas remains in the disk. Thus, our simulations represent planetesimal or debris disks that are gas-poor and dynamically cold (or at least as dynamically cold as they can be, given the presence of the planet). Given that dust subject to drag forces was found by \cite{Wyatt05} to be quickly removed, we will neglect such particles and set $\beta$, the ratio of the force due to radiation pressure to the force of gravity, to zero, as justified by our discussion in Section \ref{Sec:PRvscoll}.

The simulations are run for 1 million yr. This is found empirically to be more than long enough for MMR gaps to open and for the disks to reach a quasi-steady state; indeed, the formation of MMR gaps in our simulations typically becomes apparent in only a few orbital timescales of the planet, and the gaps are already well established by $\sim 10^4$ yr. Timesteps of approximately 25 and 50 days are chosen when the planet is exterior and interior to the disk, respectively. The timestep is adjusted slightly at the beginning of the simulation so that the planet will be at apocenter at the end of the simulation for reasons of convenience that will be discussed in Section \ref{Sec:Results}.

We measure the width and location of each MMR gap by fitting a Gaussian function to where the gap appears in a histogram of the particle distribution in the disk, as will be discussed in Section \ref{Sec:Results}. The uncertainties in the MMR gap widths and locations are calculated in the same way as in \textit{Paper I} and come from three independent sources: (1) goodness of fit from least-squares fitting to the histogram bins; (2) Poisson statistics of the particles in each bin, associated with the choice of bin size; and (3) the fit model. The different factors that contribute to the uncertainty in our calculations are then added in quadrature. Uncertainties due to the first two factors are generally small, particularly with our choice of the bin size (i.e., $0.006 ~a$), which is small enough to ensure that at the beginning of the simulation and before the disk is perturbed, each bin contains about $1\%$ of the total number of particles in the disk. However, given the fact that the MMR gaps we see in our simulations are not perfectly Gaussian, fitting such a model introduces uncertainty, particularly since the gaps are not necessarily symmetric about the mean. Therefore, to calculate the uncertainty in the measured values for the width and the location of each MMR gap, we perform the Gaussian fitting three times and record the median and width of the gap each time. The three ways that we do this are

(1) by normalizing the height of each bin to the lower edge of the gap, \\
(2) by normalizing the height of each bin to the higher edge of the gap, \\
(3) by applying no normalization.

The standard deviation between the three values obtained is the dominant source of uncertainty in our measurements.


\subsection{Simulated Debris Disks}
\label{Sec:SimDD}

The simulations are set up with the perturbing planet having a semimajor axis equivalent to that of Jupiter (i.e., $\sim$ 5.204 AU); however, this choice is arbitrary, since the physics involved scales with distance. As a result, our simulations are applicable to debris disks of all sizes; and so we normalize the scales by the planet's semimajor axis to have the planet at roughly unit distance. We vary the mass and eccentricity of the planet over a range of values at the extremes of which the disk is largely destroyed and/or the resonance gaps are completely eroded. Note that the planet is always placed either interior or exterior to the debris disk, as our interest here is not in the gap the planet clears about its orbit but rather in the structures that appear away from the orbit of the planet itself.

We place 10,000 particles per 1 AU of the disk's radial extent to be consistent with our previous work and choose the location of the inner and outer edges of the disk such that the three resonances being studied — i.e., 2:1, 3:2, and 3:1 — fall in the disk ($1.204~AU<a^\prime_D<4.204~AU$ for the interior resonance and $6.204~AU<a^\prime_D<12.204~AU$ for the exterior case). Note that the inner edge of the disk is extended 1 AU further inward for the interior resonance case compared to \textit{Paper I} for the purpose of catching higher-order resonance structures as the planet's eccentricity is increased (see Section \ref{Sec:HigherMMR}). Also, we set initial particle eccentricities to the value of the eccentricity induced by the planet at each particle's semimajor axis, i.e., the forced eccentricity.


\section{Results}
\label{Sec:Results}

The simulation results allow an examination of the spatial distribution of the disk and any resulting structures. Unlike our solar system's main asteroid belt, where an externally taken telescopic image would not reveal the Kirkwood gaps due to particle eccentricities smearing them out, the appearance and the size of the MMR gaps in our simulated disks suggest that such structures would likely be observable in telescopic images of some quiescent debris disks, as will be discussed in Section \ref{Sec:ALMA_Sim}.

In order to study structures formed by MMRs, histograms of the number distribution of disk particles in heliocentric distance for different quadrants of the disk are made. The MMR gaps can be fit by a Gaussian function as a first approximation for comparison with the analytic measurements. Observers often measure gap widths by locating where the disk brightness drops to half the peak value around a gap \citep[see, for instance,][]{Chiang09}. We choose the range for our Gaussian fitting in the same way here; however, it is often challenging to define the edges of the gaps in our simulations, as the particle distribution is not smooth. The difficulty in defining the edges of the gaps and their asymmetries introduces an uncertainty in our calculations that is addressed in Section \ref{Sec:Method}.

In \textit{Paper I}, it was shown that MMR gaps are often azimuthally asymmetric. Therefore, we again divide the disk into four equal segments, this time about the line of apses of the planet's orbit. This coincides with the line of apses of the disk particles due to their pericenter alignment with the planet (see Section \ref{Sec:PeriAlign_Offset}). We then fit Gaussian functions to where the gaps appear and compare the width and location of each MMR gap to the analytical values found using the equations discussed in Section \ref{Sec:Intro}. Also provided are expressions for calculating the mass and some orbital parameters of the perturbing planet even if the planet itself is not resolved in the observations. Therefore, our technique is an indirect method to detect and characterize extrasolar planets in systems with debris disks.

One key finding of \textit{Paper I} was that a slight increase in the eccentricity of the planet (i.e., when $e=e_J=0.0489$, where the subscript $J$ refers to Jupiter) resulted in an extra gap appearing in the disk that opened at the 3:1 MMR with the planet. That work is extended here by varying the planet eccentricity to much higher values. We find that this gap becomes more prominent and that other higher-order resonances also appear in the disk as the planet eccentricity is increased further. Furthermore, we saw in \textit{Paper I} that the width of an MMR gap is related to the perturber's mass, and here again we find that the properties of the gaps allow us to constrain the mass of the planet. 

The eccentricity of the disks examined here means that one side is narrower by $a^\prime_{D_o} e^\prime_{D_o} - a^\prime_{D_i} e^\prime_{D_i}$ at the pericenter side and wider by the same amount at the apocenter, where $a^\prime_D$ and $e^\prime_D$ are the semimajor axis and eccentricity of the disk, respectively, with the subscripts $o$ and $i$ denoting the outer and inner disk edges. Therefore, the theoretical locations and widths of MMR gaps in each segment need to be adjusted by $\frac{1-e_f^{\prime^2}}{1+e_f^\prime \cos(\nu^\prime)}$, where $\nu^\prime$ is the true anomaly of the particles' orbits at the center of each segment and is taken to be $0\degr$ and $180\degr$ in the two segments whose centers lie on the line of apses at pericenter and apocenter, respectively, and $90\degr$ and $-90\degr$ in the other two segments.

At the same time, increasing the mass of the perturber causes the disk edge to erode due to gravitational scattering by the more massive planet. The combination of the two effects, i.e., large perturber mass and large perturber eccentricity, leads us to expect the resonance structures to be eventually destroyed. Our initial planetary orbital eccentricity is set at 0.1 and is increased by increments of $0.05$. The mass is increased by increments of $1.0 ~M_J$, starting with a $1~ M_J$ planet, until the first- and second-order resonance gaps can no longer be observed. It must be noted that our earlier work (i.e., \textit{Paper I}) shows that MMR gaps can theoretically be opened by much less massive planets of the order of 1 Earth mass, though these gaps would be very narrow. Thus, we choose to focus on planets with $M > 1.0~ M_J$ in this study, since they would be relatively more likely to be revealed in high-resolution telescopic images of debris disks. The results are discussed in Sections \ref{Sec:IntRes} and \ref{Sec:ExtRes} for interior and exterior MMRs, respectively.


\subsection{Interior Resonance}
\label{Sec:IntRes}

If a resonance is caused by the more massive object having a larger orbit than the orbit of the object it is perturbing, it is referred to as "interior resonance." MMRs caused by Jupiter on the asteroid belt are an example of this type of resonance.

In \textit{Paper I}, we noted that not only is the 2:1 interior MMR with a disk of debris material not azimuthally symmetric about the star, it forms two arc-shaped gaps whose centers are at the planet's inferior conjunction and superior conjunction, both of which orbit the star at the same rate as the planet itself. 

\begin{figure*}
    \centering
    \subfigure[]{%
    \includegraphics[totalheight=0.25\textheight]{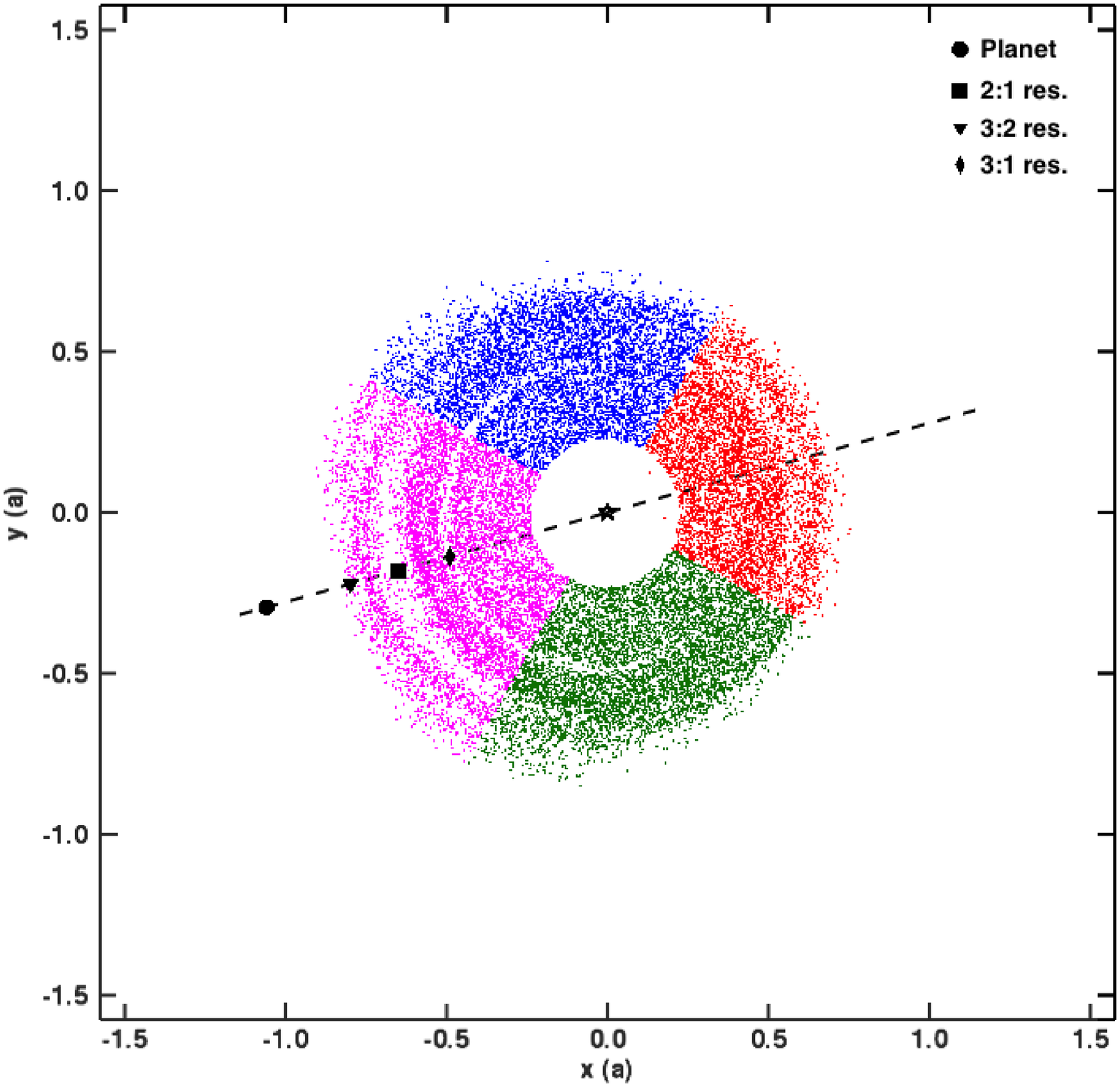}
    \label{Fig:outerE0.1m3-disk}}
\hfill
    \subfigure[]{%
    \includegraphics[totalheight=0.45\textheight]{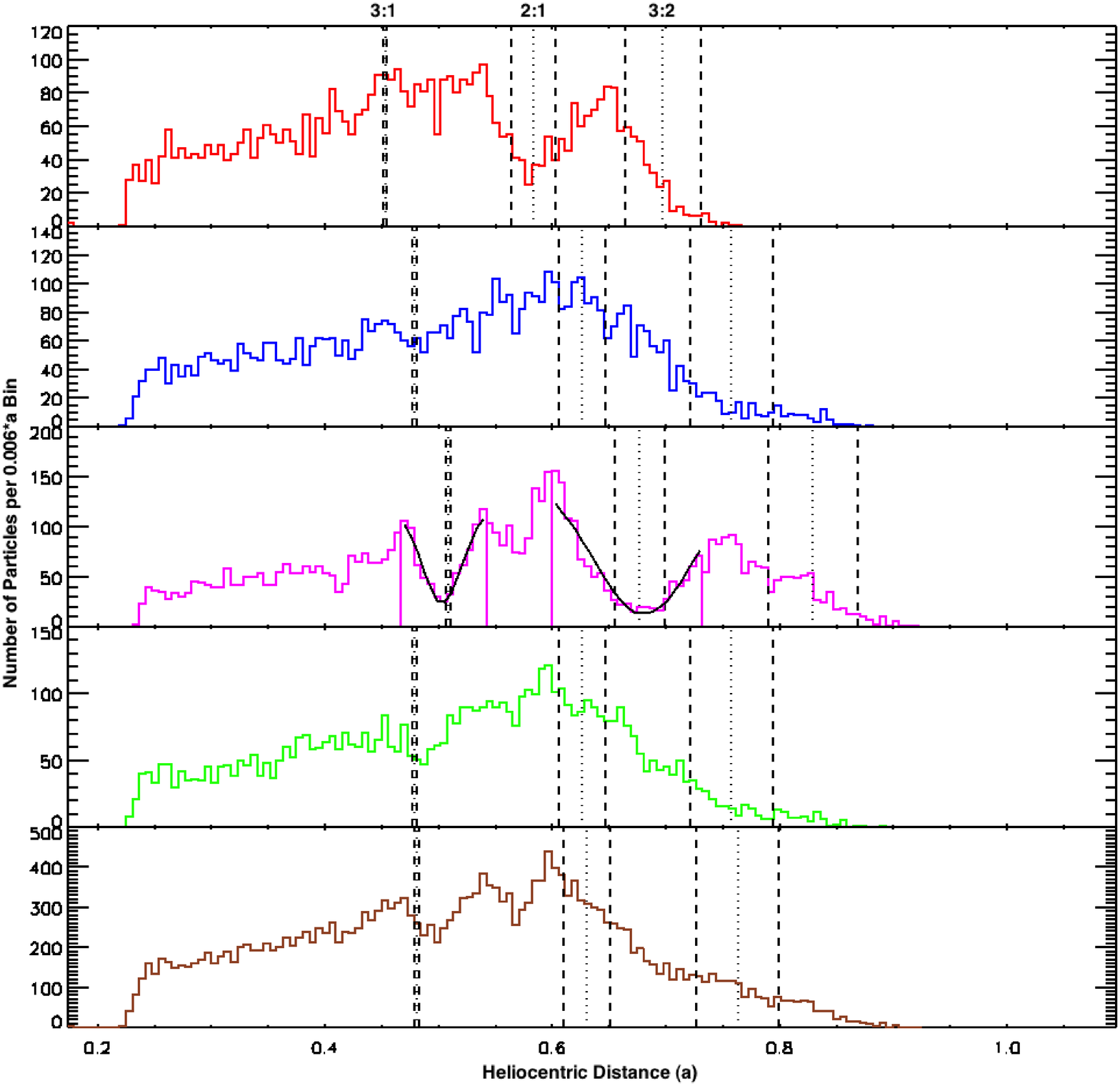}
    \label{Fig:outerE0.1m3-hists}}
\caption{Simulation result illustrating interior MMRs with a single planet of mass $M=3.0 ~M_J$ and eccentricity $e=0.1$. (a) The interior 2:1 resonance with a single planet (filled circle) forms two arc-shaped gaps in the disk whose centers trace the planet as it orbits the star, while the 3:1 MMR gap appears as a single arc of less width and is fixed at the disk's apocenter. The dashed line is the planet's line of apses. Note the alignment of the disk's line of apses with the planet's. The symbols represent the theoretical (or nominal) locations for the 2:1, 3:2, and 3:1 MMRs. (b) Distribution of disk particles in heliocentric distance. The colors in the top four panels correspond to the segments of the same color in Figure \ref{Fig:outerE0.1m3-disk}, while the bottom panel represents the overall distribution of the disk particles from the four segments put together. The dotted lines are the nominal locations of each gap at the 2:1, 3:2, and 3:1 interior resonance with the planet found through Equation \ref{Eq:MMR_Def_a}, while the dashed lines show the width of each gap calculated analytically. For consistency, we chose the same bin size as that in \textit{Paper I} which is $ 0.006 ~a $, where $a$ is the planet's semimajor axis. Gaussian fits are made to both gaps in the middle histogram, which corresponds to the region close to the planet's (and hence the disk's) apocenter.}
\label{Fig:outerE0.1m3-disk-hists}
\end{figure*}

Here we see the same double-arc feature at the 2:1 MMR at higher planet eccentricity values (see Figure \ref{Fig:outerE0.1m3-disk}). Furthermore, the additional gap at 3:1 again appears when $e>0$ and becomes more prominent as the eccentricity is increased further. This 3:1 MMR gap forms a single arc at the disk's apocenter. Unlike the 2:1 MMR, which travels around the disk at the same rate as the planet, the 3:1 resonance gap remains at the apocenter, regardless of where the planet is along its orbit. This difference can help localize the planet if it does not appear in a telescopic image of a real disk system. Thus, the two resonances together provide information on planetary position and orbit geometry.

The appearance of the 3:1 gap at apocenter makes it more visible than it would be if it were at pericenter. The disk is thicker near the forced apocenter, and the MMR gaps that are in this part of the disk are wider by a factor of $1+e_f^\prime$. This is not the case for exterior resonances, as will be discussed in Section \ref{Sec:ExtRes}.

Since the 3:1 gap remains fixed at the disk's apocenter while the 2:1 MMR moves with the planet, for clarity, the simulation time is modified slightly so that the planet finishes at apocenter when the simulation ends. An example is shown in Figure \ref{Fig:outerE0.1m3-disk-hists} for MMRs with a planet with $M=3.0 ~M_J$ and $e=0.1$. The theoretical locations of the 2:1, 3:2, and 3:1 interior MMRs are shown by symbols in Figure \ref{Fig:outerE0.1m3-disk} and by dotted lines in Figure \ref{Fig:outerE0.1m3-hists} and are calculated using Equation \ref{Eq:MMR_Def_a}. The dashed lines on the histograms show the theoretical width of each gap found by the equations for $\delta a^{\prime}_{max}$.

\begin{figure*}
    \centering
    \includegraphics[totalheight=0.2\textheight]{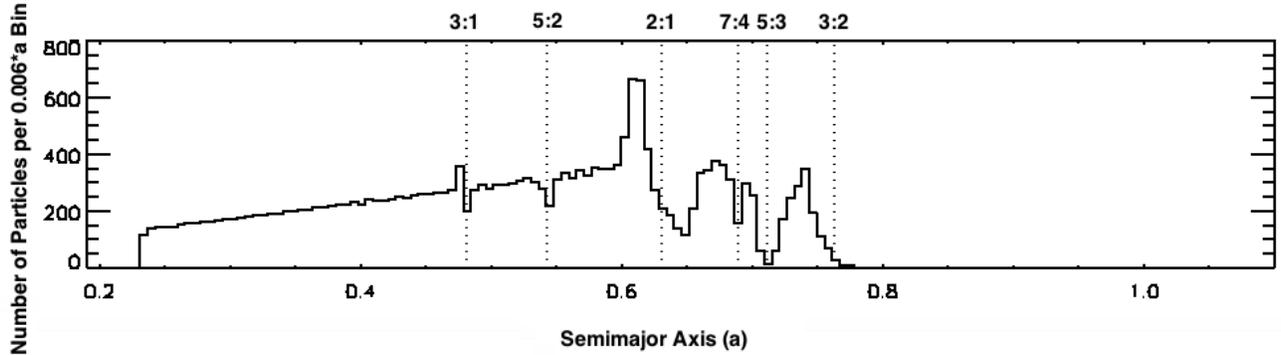}
    \caption{Distribution of particles shown by Figure \ref{Fig:outerE0.1m3-disk-hists} but in the semimajor axis instead of heliocentric distance. The dotted lines show the nominal resonance locations of the gaps. We note the appearance of additional gaps in this figure at 5:2, 7:4, and 5:3 MMR with the planet that are not revealed when particles are plotted in heliocentric distance. We also note a pileup of particles inward of the 2:1 gap, but we have not explored how to disentangle particles near MMRs being removed versus shifting their positions (though both result in gap formation in the disk).}
    \label{Fig:Na_a}
\end{figure*}

As noted earlier, the presence of MMR gaps does not necessarily mean that the particles have been ejected from their orbits or destroyed by collision with the star or a planet. They are often caused by an increase in the particles' orbital eccentricities that shifts them away from the heliocentric distance in question. Whether a particle gets physically removed by resonant interaction with a planet or stays bound to the system but increases its eccentricity can be examined by plotting particle distributions in the semimajor axis. This is shown by Figure \ref{Fig:Na_a}, which is the same as the bottom panel of Figure \ref{Fig:outerE0.1m3-hists} but plotted in the semimajor axis instead of heliocentric distance. Note that in our simulations, particles are removed both when they go into a hyperbolic orbit \textit{or} when they go beyond $0.05 ~AU < r_p^\prime < 1000 ~AU$, where $r_p^\prime$ is the particle-star distance. So the gaps in Figure \ref{Fig:outerE0.1m3-hists} are particles being shifted in their orbits, as well as removed outright. Plotting particle distribution in the semimajor axis reveals a few other MMR gaps in addition to the ones that we study here. In fact, the solar system's Kirkwood gaps only appear when asteroid distribution is plotted in the semimajor axis. This is because gaps can be smeared out due to particle eccentricities that can bring them in and out of the gaps (see Figure 1 in \cite{Tabeshian16} for a plot of main-belt asteroid distribution in heliocentric distance). However, here we argue that this may not necessarily be the case for all debris disks, and a telescopic image of these disks may reveal MMR gaps.

\begin{figure*}
    \centering
    \subfigure[]{%
    \includegraphics[totalheight=0.25\textheight]{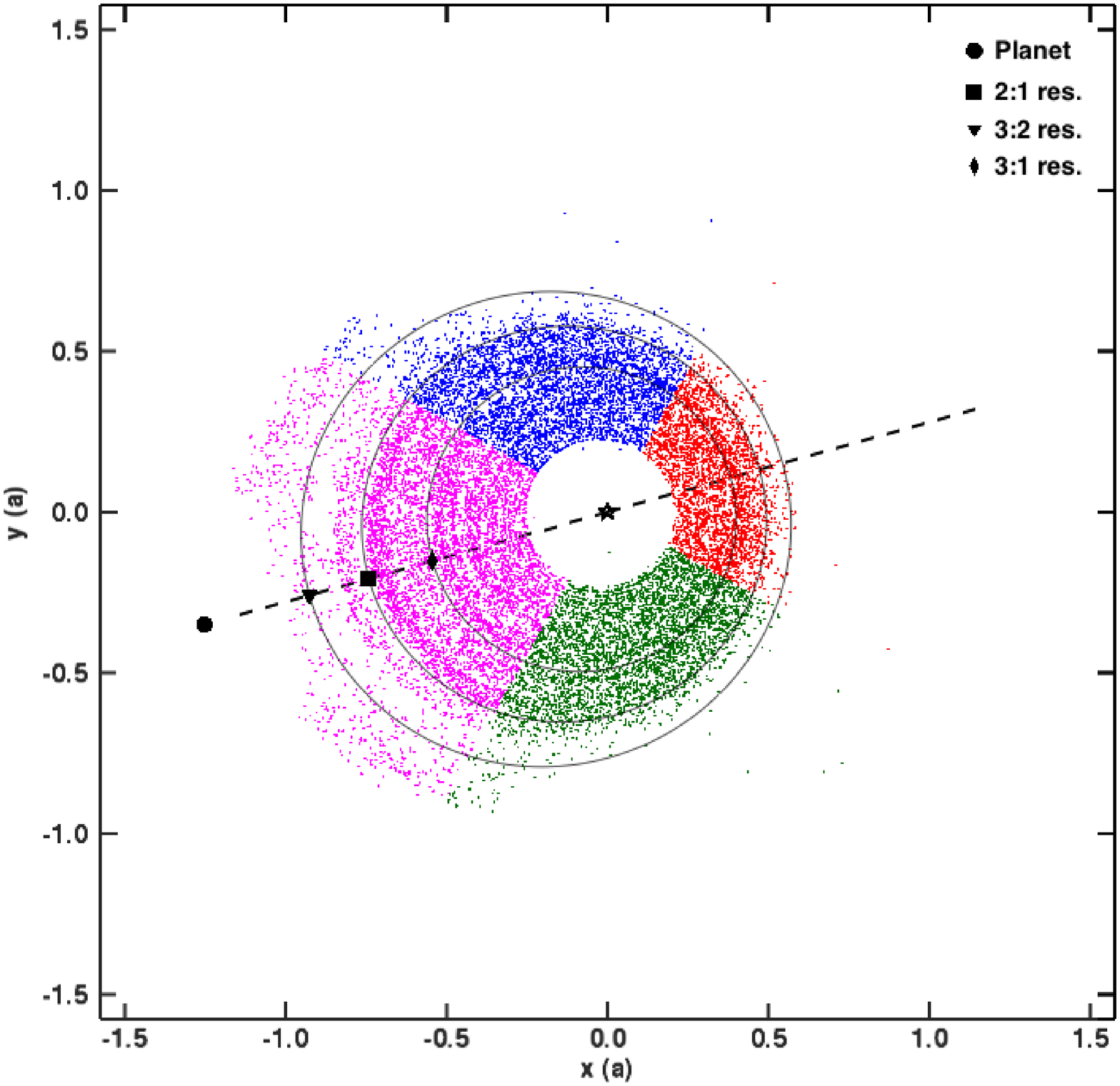}
    \label{Fig:outerE0.3m1-disk}}
\hfill
    \subfigure[]{%
    \includegraphics[totalheight=0.45\textheight]{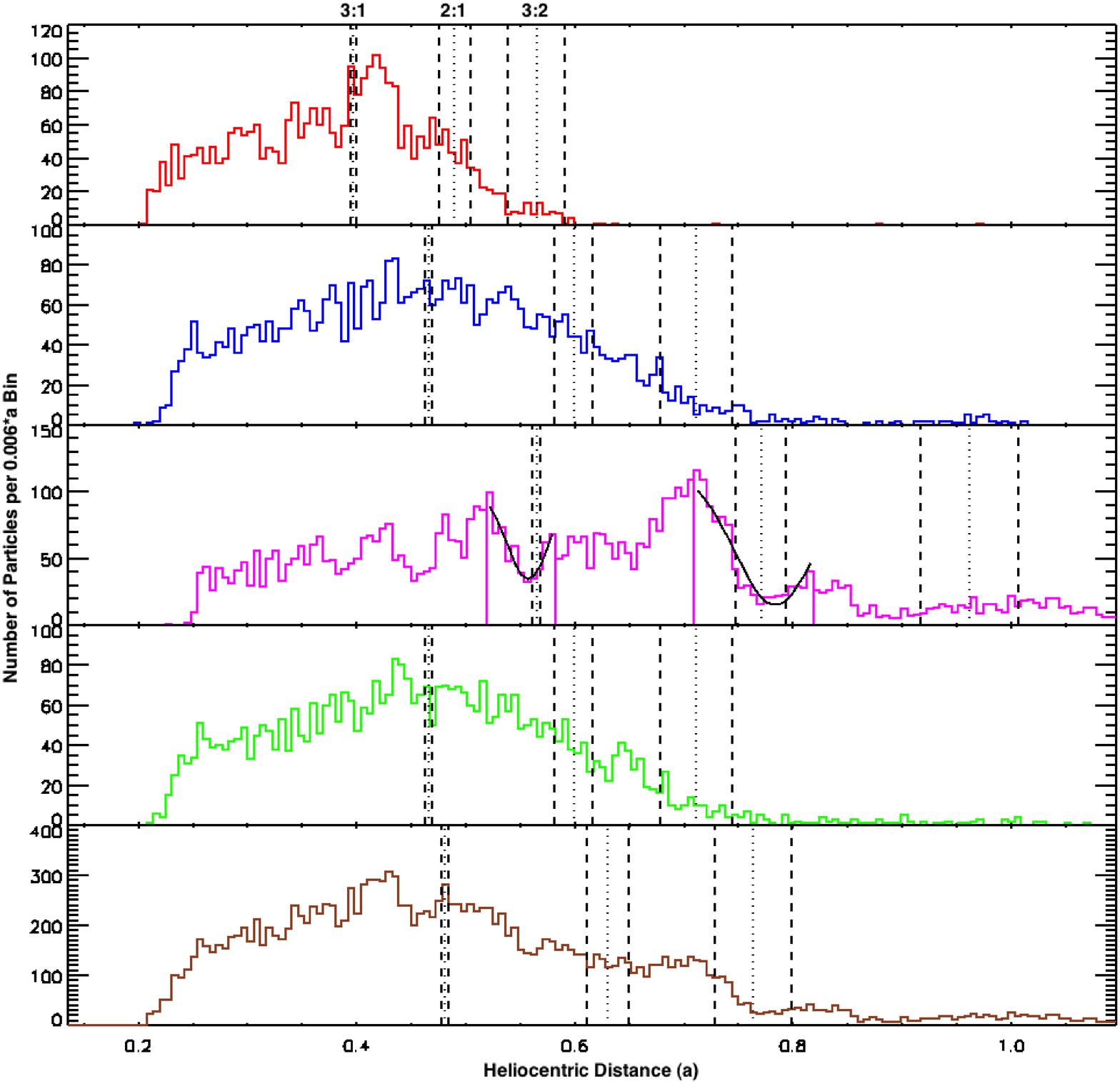}
    \label{Fig:outerE0.3m1-hists}}
\caption{Interior resonance gaps formed by the interaction of disk particles with a planet of mass $M=1.0 ~M_J$ and $e=0.3$. (a) Ellipses are drawn on the disk to show the theoretical locations of resonances in all segments. Note that the 2:1 MMR gap is almost at the edge of the disk in the red region (pericenter), while the same gap is more evident in the region near apocenter (magenta region). There is particle trapping at 3:2 MMR with the planet. (b) Histograms showing particle distribution for each segment marked with the same colors as in panel (a).}
\label{Fig:outerE0.3m1-disk-hists}
\end{figure*}

Increasing the planet's eccentricity eventually erodes the disk and destroys the 2:1 arc there. This is illustrated in Figure \ref{Fig:outerE0.3m1-disk-hists} for MMRs with a planet with $M=1.0 ~M_J$ and $e=0.3$.

In some cases, particles persist outside the main disk, ahead of and behind the planet, as seen in Figure \ref{Fig:outerE0.3m1-disk-hists}. These are analogous to the Hilda asteroids in the solar system's main asteroid belt, which are trapped in 3:2 MMR with Jupiter and form a triangular-shaped pattern with their apexes fixed relative to Jupiter. We see this in our simulations of low-mass planets with moderate orbital eccentricities.

\begin{figure*}
    \centering
    \subfigure[]{%
    \includegraphics[totalheight=0.25\textheight]{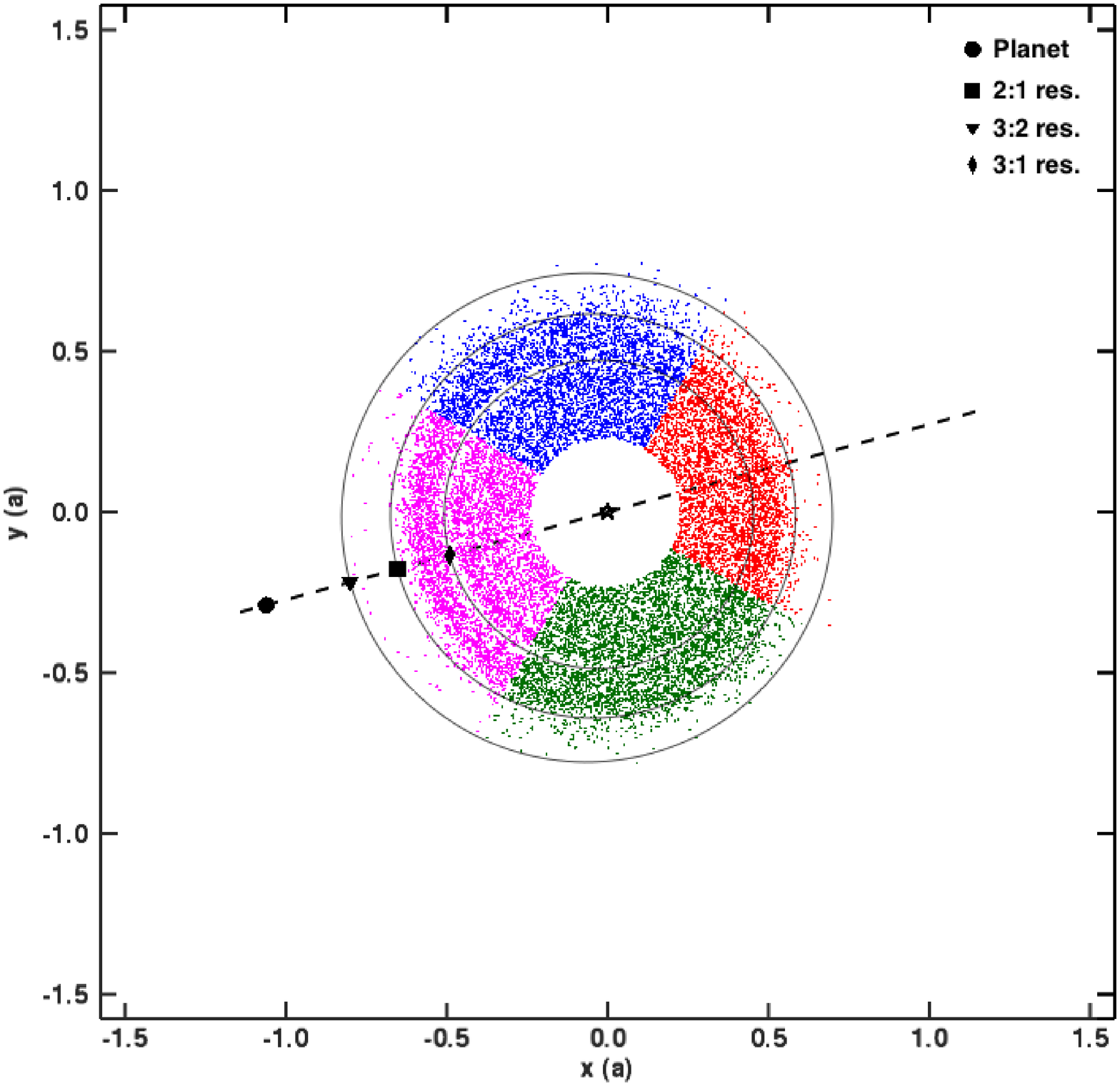}
    \label{Fig:outerE0.1m6-disk}}
\hfill
    \subfigure[]{%
    \includegraphics[totalheight=0.45\textheight]{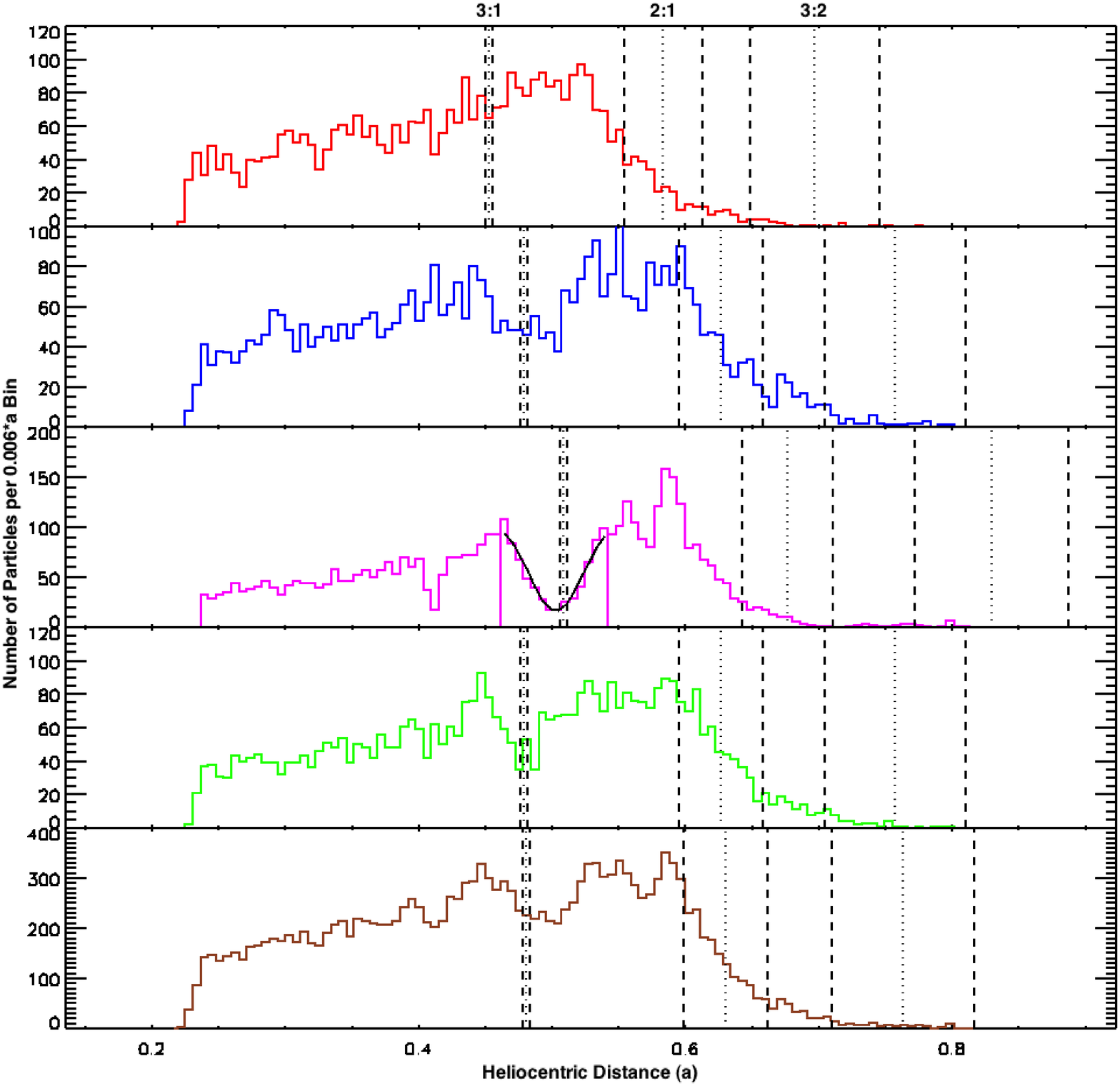}
    \label{Fig:outerE0.1m6-hists}}
\caption{Same as Figure \ref{Fig:outerE0.3m1-disk-hists} but with $M=6.0 ~M_J$ and $e=0.1$. Note that the 2:1 gap is destroyed on both sides of the disk. This happens when the perturber's eccentricity is low but its mass, and hence its Hill radius, is increased. The most prominent gap in this case is the 3:1, though the 4:1 also now becomes distinguishable (see Section \ref{Sec:HigherMMR} and Figure \ref{Fig:outerE0.2m6-disk-hists} for a discussion of higher order resonances).}
\label{Fig:outerE0.1m6-disk-hists}
\end{figure*}

While increasing planet eccentricity makes the disk narrower at its pericenter, increasing its mass erodes it on both sides (i.e., near apocenter and pericenter), as seen in Figure \ref{Fig:outerE0.1m6-disk-hists}, where $M=6.0 ~M_J$ and $e=0.1$.

\subsection{Exterior Resonance}
\label{Sec:ExtRes}

An exterior resonance occurs when the more massive object perturbs the orbit of an object exterior to its orbit. One key result in \textit{Paper I} was that, whereas the 2:1 interior MMR forms two arc-shaped gaps, the gap formed at the 2:1 exterior resonance with a planet is a single arc at the perturber's opposition. This difference allows one to distinguish interior from exterior resonance even if the planet that is causing it remains unseen. 

In \textit{Paper I}, increasing the orbital eccentricity of the planet to $\sim 0.05$ resulted in the 2:1 exterior gap being extended azimuthally, while at the same time a second gap appeared at a location corresponding to the 3:1 exterior resonance with the planet. Furthermore, for low orbital eccentricities, we observed what seemed to be a series of tightly wound spiral waves originating from the 3:1 MMR that we interpreted as forced eccentricity waves due to Lindblad resonances, similar to what is seen in Saturn's rings. Here we report that these waves become weaker and eventually disappear when the planet's orbital eccentricity is increased beyond $0.2$ or when $m > 2.0 ~M_J$. 

\begin{figure*}
    \centering
    \subfigure[]{%
    \includegraphics[totalheight=0.24\textheight]{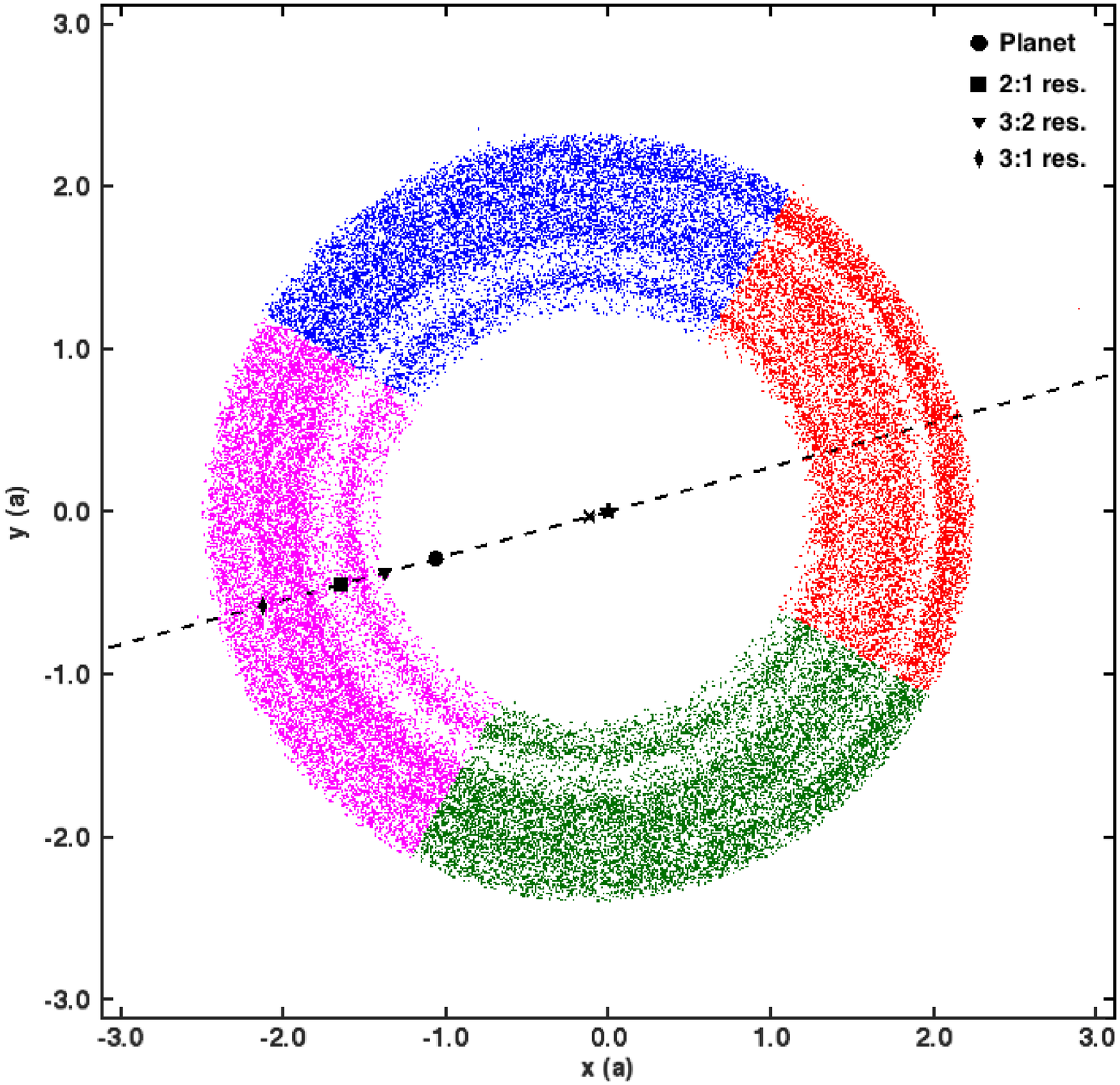}
    \label{Fig:innerE0.1m3-disk}}
\hfill
    \subfigure[]{%
    \includegraphics[totalheight=0.45\textheight]{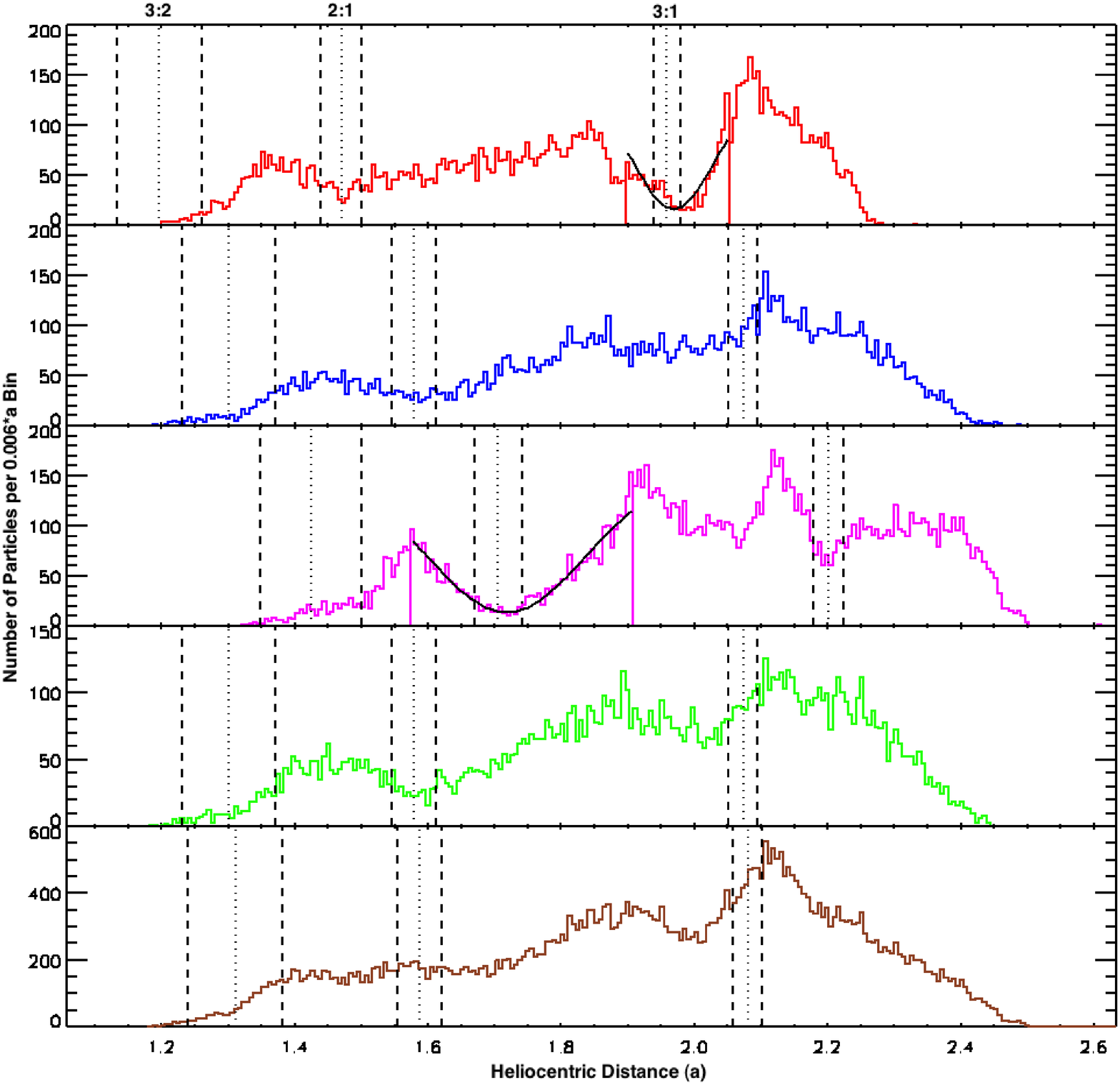}
    \label{Fig:innerE0.1m3-hists}}
\caption{Same as Figure \ref{Fig:outerE0.1m3-disk-hists}, with mass $M=3.0 ~M_J$ and eccentricity $e=0.1$, but with the perturber interior to the disk. (a) The exterior gap at the 2:1 MMR with a single planet forms a single arc in the part of the disk closest to the planet, while its center tracks the planet. The 3:1 MMR also appears as a single arc but has a smaller width and is fixed at the disk's pericenter. The cross shows the geometric center of the disk, taken to be midway between the inner edges along the major axis. (b) Particle distribution for each segment of the same colors as in panel (a), along with Gaussian fits to the two gaps.}
\label{Fig:innerE0.1m3-disk-hists}
\end{figure*}

The gap at the 3:1 exterior MMR also becomes wider as the planet eccentricity is increased, more so if the mass of the planet is increased along with its eccentricity. The behavior of this gap is similar to that in the 3:1 interior MMR in that neither of them move around the disk in the inertial frame. However, whereas the latter is found to be fixed at the disk's apocenter, the exterior MMR gap at 3:1 is fixed at the disk's pericenter. Resonance structures formed at the 2:1 and 3:1 exterior resonances are shown in Figure \ref{Fig:innerE0.1m3-disk-hists}.

\section{Discussion}
\label{Sec:Discussions}

The main results of this work are twofold. First, if a planet is observed near a debris disk, the planet's semimajor axis, orbital eccentricity, and mass can in principle be determined from its resonant features within the disk. Second, the presence of unseen planets can be inferred from resonant structures within the disk, but more importantly, (1) the planet's position can be narrowed down for more efficient targeted searches, and (2) the planet's properties can still be determined almost as easily as if the planet itself had been detected. Because it is typically easier to detect the debris disk structures than the planet itself, and since the planet does not need to be detected for these measurements to be made, below we discuss an algorithm for the determination of the properties of the planet on the assumption that the resonant structures have been observed at a single epoch but that the planet itself is unseen. Where relevant, shortcuts available when multi-epoch observations are available will be outlined.

First, we discuss in sections \ref{Sec:Asymm} and \ref{Sec:HigherMMR} how the appearance and shape of these gaps alone can reveal some information about an unseen planet.


\subsection{Locating an Unseen Planet from Asymmetries in MMR Gaps}
\label{Sec:Asymm}

The azimuthal asymmetry, as well as the difference in the physical appearance of MMR gaps for interior and exterior resonances, can be used to not only distinguish resonance gaps from those formed by the dynamical clearing of a planet's orbit but also to determine on which side of the disk the perturber lies if it is unseen.

Gaps cleared out by planets sweeping up their surroundings are azimuthally symmetric and have been observed in both protoplanetary disks \citep[e.g., HL Tau; see, for instance,][]{Brogan15} as well as in debris disks \citep[e.g., Epsilon Eridani; see][]{Backman09}. However, based on our simulations, MMR gaps do not show azimuthal symmetry. As described in Sections \ref{Sec:IntRes} and \ref{Sec:ExtRes}, the 2:1 gap forms two arcs whose centers are at the planet's inferior and superior conjunctions when the planet is placed exterior to the disk (i.e., when $a > a^\prime$) while forming a single arc with its center at the planet's opposition when the perturber is interior to the disk (i.e., when $a < a^\prime$). The different shapes of resonant structures can be understood through geometrical arguments and are discussed in Chapter 8 of \cite{Murray99}. Nevertheless, the two distinctive gap shapes observed in our simulations for interior and exterior resonances suggest that even if the planet's location is unknown, the appearance of either of these gaps can be used to direct targeted searches to locate the planet.

Moreover, if there is an additional arc-shaped gap in the disk, it is likely to be the 3:1 resonance structure if the following three conditions are met: (1) one gap is narrower than the other; (2) the narrower gap is interior to the double-arc gap and the ratio of their locations is approximately  $(\frac{3}{2})^{-2/3} = 0.8$, or it is exterior to the wider single-arc gap with the ratio of their locations being about $(\frac{3}{2})^{2/3} = 1.3$; and (3) the narrower gap is fixed at either the apocenter or the pericenter of the disk, while the other gap orbits the star. The last condition requires multi-epoch observation of the disk; nevertheless, even if the disk is only observed once, we can still distinguish the 2:1 from the 3:1 gap based on the other two conditions. 

It must be noted, however, that gaps that are formed by dynamical clearing of planets in their surroundings can become asymmetric, forming a horseshoe structure at the planet's orbit if there is a substantial number of particles trapped in the planet's 1:1 MMR, such as the so-called Trojan asteroids. This could make it more challenging to distinguish an MMR gap from that formed in the feeding zone of a planet, unless there are additional resonance gaps, which may be used to identify resonance locations as discussed above.


\subsection{Higher-order Resonances}
\label{Sec:HigherMMR}

\begin{figure*}
    \centering
    \subfigure[]{%
    \includegraphics[totalheight=0.25\textheight]{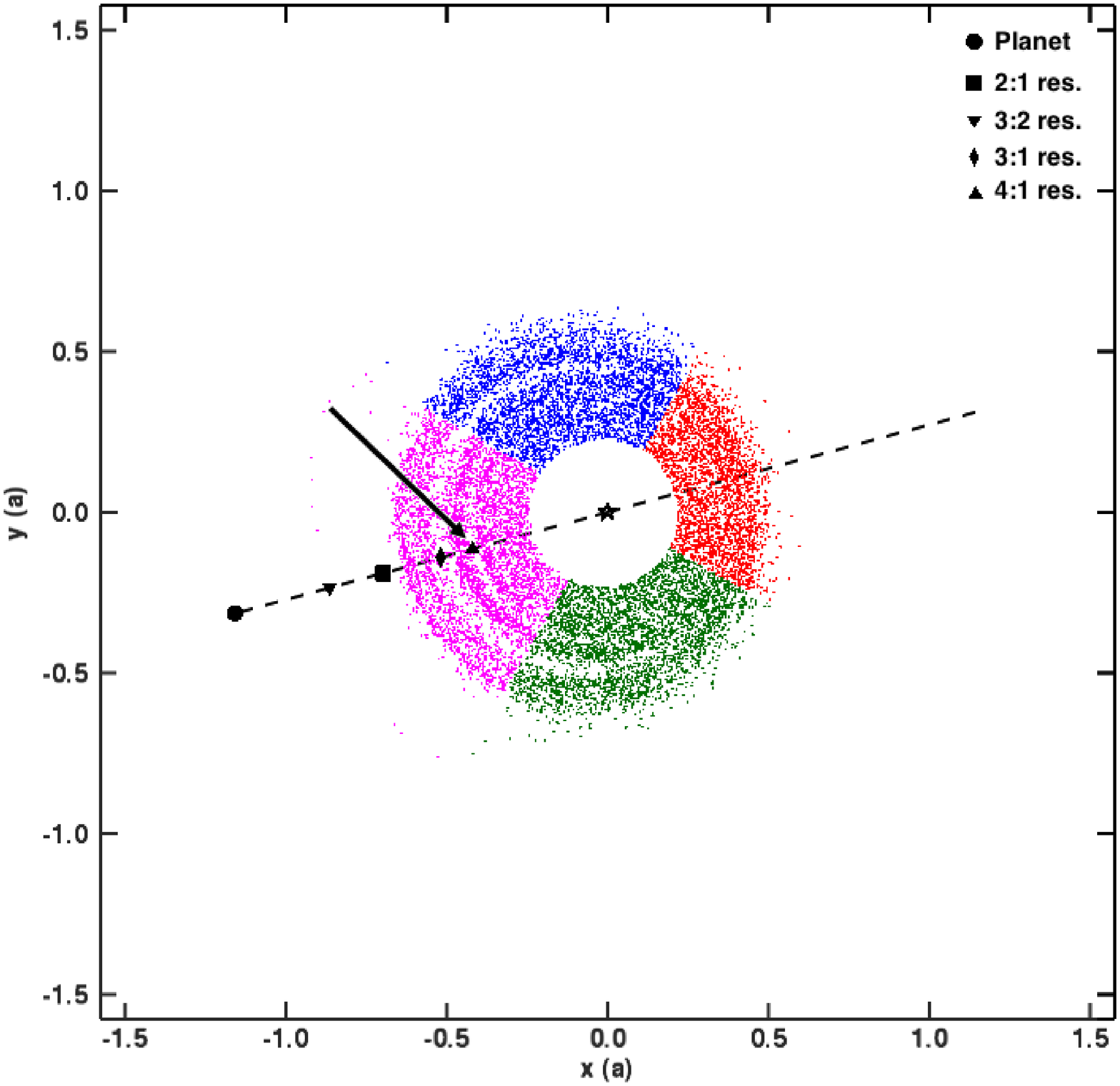}
    \label{Fig:outerE0.2m6-disk}}
\hfill
    \subfigure[]{%
    \includegraphics[totalheight=0.45\textheight]{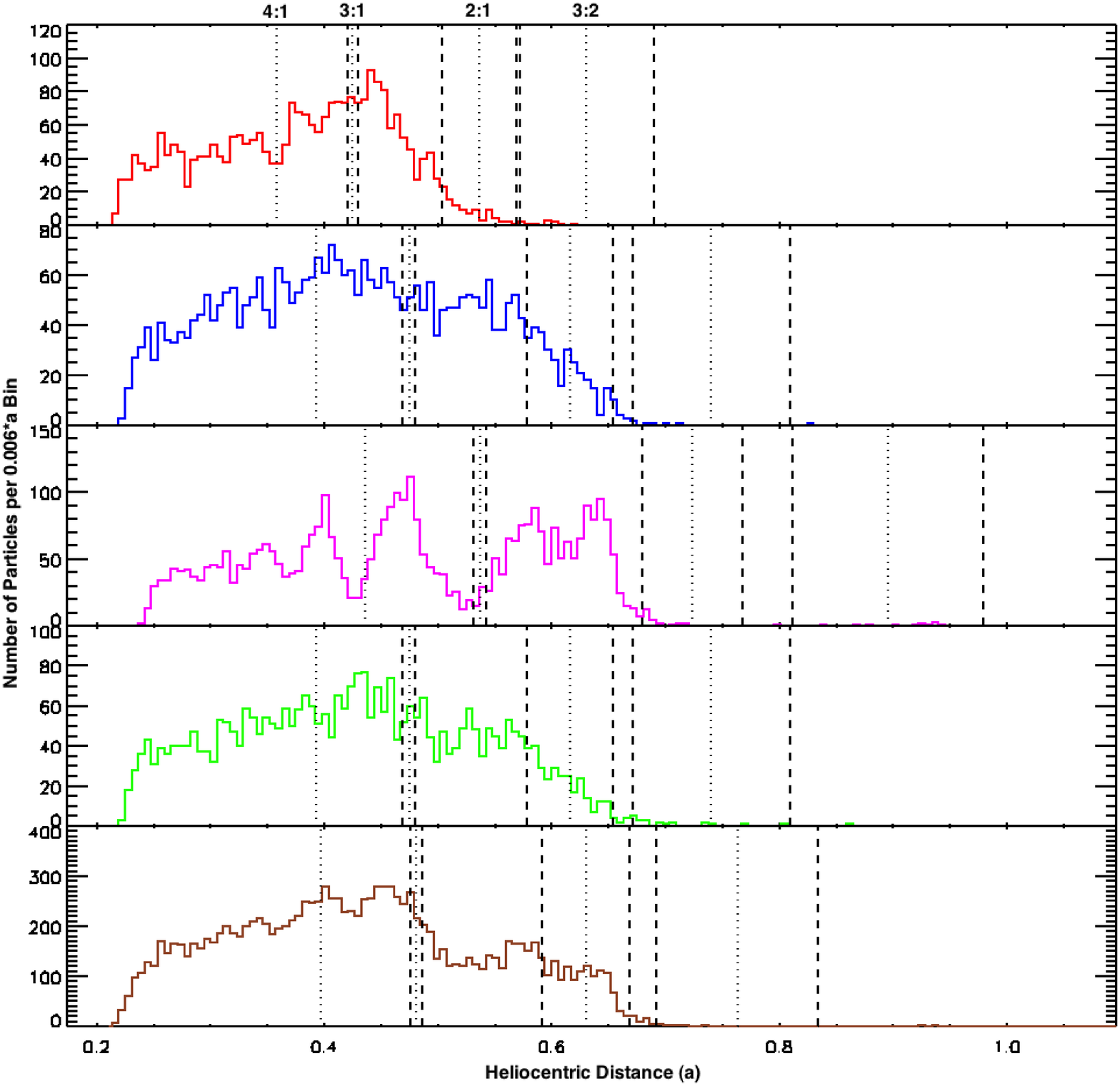}
    \label{Fig:outerE0.2m6-hists}}
\caption{Interior MMR with a planet of mass $6 ~M_J$ and eccentricity $0.2$. (a) Increasing the perturber eccentricity results in formation of an additional gap, corresponding to the 4:1 interior MMR and shown with an upright triangle and an arrow pointing to it, which becomes wider as the perturber's mass is increased along with its eccentricity. (b) The theoretical location of the 4:1 gap is also overplotted on the histograms.}
\label{Fig:outerE0.2m6-disk-hists}
\end{figure*}

As the planet's eccentricity is increased, the 2:1 gap disperses first as the disk around it is scattered by the planet, followed by the destruction of the 3:1 gap if the planet's eccentricity is increased further. Nevertheless, though the 2:1 and 3:1 MMR gaps get eroded as the orbital eccentricity of the planet is increased, higher-order resonances start appearing in the disk even before the other two gaps disappear, particularly at higher planet masses. This is shown by Figure \ref{Fig:outerE0.2m6-disk-hists}, where the mass and eccentricity of the planet are $6 ~M_J$ and $0.2$, respectively. The extra gap seen in this figure is at 4:1 interior resonance with the planet. This gap already appears at lower planet mass-eccentricity combinations, such as when $M = 3.0 ~M_J$ and $e = 0.1$ or $M = 1.0 ~M_J$ and $e = 0.15$, and becomes more prominent as the two parameters are increased.

\subsection{Determining Planetary Parameters}
\label{Sec:Params}

In this section, we discuss how MMR gaps can be used to obtain the perturbing planet's orbital eccentricity and semimajor axis and hence its orbital period, as well as its mass. Thus, observers can detect and characterize extrasolar planets based on their resonant effects on debris disks even if the planets themselves have yet to be detected directly or through other methods.


\subsubsection{Eccentricity}
\label{Sec:e_pl}

In order to determine the orbital eccentricity, $e$, of a planet that perturbs a debris disk, an ellipse should be fitted to the gap in order to calculate its eccentricity. The gap eccentricity is equal to the forced eccentricity, $e_f^{\prime}$, induced by the planet; so we can use Equation \ref{Eq:Forced_e} to calculate the eccentricity of the perturber. The advantage of using this equation for the measurement of planet eccentricity based on the eccentricity of an MMR gap is that the semimajor axis of the gap or the planet need not be known as long as the particular resonance gap can be identified. This is due to the fact that Equation \ref{Eq:Forced_e} only requires the \textit{ratio} of the two semimajor axes, which can be found from p and q through Equation \ref{Eq:MMR_Def_a}, without the need for individual parameters to be first determined. If both the 2:1 and the 3:1 gaps are observed, this may be established with confidence. If not, then a single gap is most likely the 2:1, the most prominent, unless the disk is heavily eroded due to large planet mass and/or eccentricity, as discussed in Section \ref{Sec:HigherMMR}.


\subsubsection{Semimajor Axis}
\label{Sec:a_pl}

Calculating the semimajor axis, $a$, of an unseen planet from a single-epoch observation of MMR gaps requires the distance to a debris disk to be known; we will assume this is at least approximately known here. Again, the MMR gap must be identified, and the $p$ and $q$ should be known. The angular separation between a gap seen in the disk to the star can be used to determine the gap distance from the star, $r^\prime$. If the disk has nonzero eccentricity, its center of symmetry is offset from the star away from the disk's pericenter. Using the equation for an ellipse in polar coordinates, shown by Equation \ref{Eq:Ellipse}, one can determine the semimajor axis of the gap's center, $a^\prime$,

\begin{equation}
\label{Eq:Ellipse}
r^\prime = \frac{a^\prime (1-e_f^{\prime^2})}{1+e_f^\prime \cos(\nu^\prime)} ~,
\end{equation}

\noindent where $e_f^\prime$ is the forced eccentricity of the gap found as described in Section \ref{Sec:e_pl}. Again, $\nu^\prime$ is the true anomaly of the gap's center and corresponds to the angle from pericenter to the gap. If no offset is observed, then $a^\prime$ is simply equal (or close to) $r^\prime$. In either case, once $a^\prime$ is determined, Equation \ref{Eq:MMR_Def_a} can be used to calculate the planet's semimajor axis, $a$, provided that we can determine which MMR gap is observed in the disk based on its shape and/or location.


\paragraph{Semimajor Axis from Multi-epoch Observations}
\label{Sec:P_pl}

As mentioned in Section \ref{Sec:Results}, we find that the 2:1 gap orbits the star at the same rate as the planet. \cite{Wyatt03} also found that patterns formed by resonant trapping of particles co-orbit with the planet, much like the pattern formed by the Hilda asteroids, which appears fixed with respect to Jupiter. Though the particles themselves are on Keplerian orbits and orbit at a rate that depends on their semimajor axis, the pattern formed by their resonant trapping goes around the star at the same rate as the planet. The same is true for patterns formed by resonant gaps at the first-order interior or exterior 2:1 resonance. This means that if multi-epoch observations of the disk are available, the orbital period of the planet, and hence its semimajor axis, can be determined by measuring the rate at which the gap moves around the disk. Of course, if the planet itself can be seen in multi-epoch observations, the orbital period is trivial to compute.

Multi-epoch ALMA observations of debris disks should soon be available. For instance, a 1.0 $M_J$ planet at 40~AU orbiting interior to the outer belt of the $\epsilon$ Eridani system, as proposed by \cite{Quillen02}, could form a gap at the 2:1 MMR, 19\arcsec ~ from the star, considering the distance to $\epsilon$ Eridani (3.22~pc). If the gap co-orbits with the planet around the 0.82 $M_\odot$ star, its orbital period will be about 280 yr, which corresponds to a motion of 0.\arcsec42~yr$^{-1}$. This means that the gap's orbital motion could be detectable in high-resolution observations within a few years. The advantage of finding the semimajor axis of the planet from its orbital period is that it eliminates the need to know the distance to the system being studied. Once the semimajor axis of the planet is known, the semimajor axis of the gap can be calculated directly from Equation \ref{Eq:MMR_Def_a} without the need to find $r^\prime$ first, provided that $p$ and $q$ are known.



\subsubsection{Mass}
\label{Sec:WvsMP}

\begin{figure*}
    \centering
    \includegraphics[totalheight=0.6\textheight]{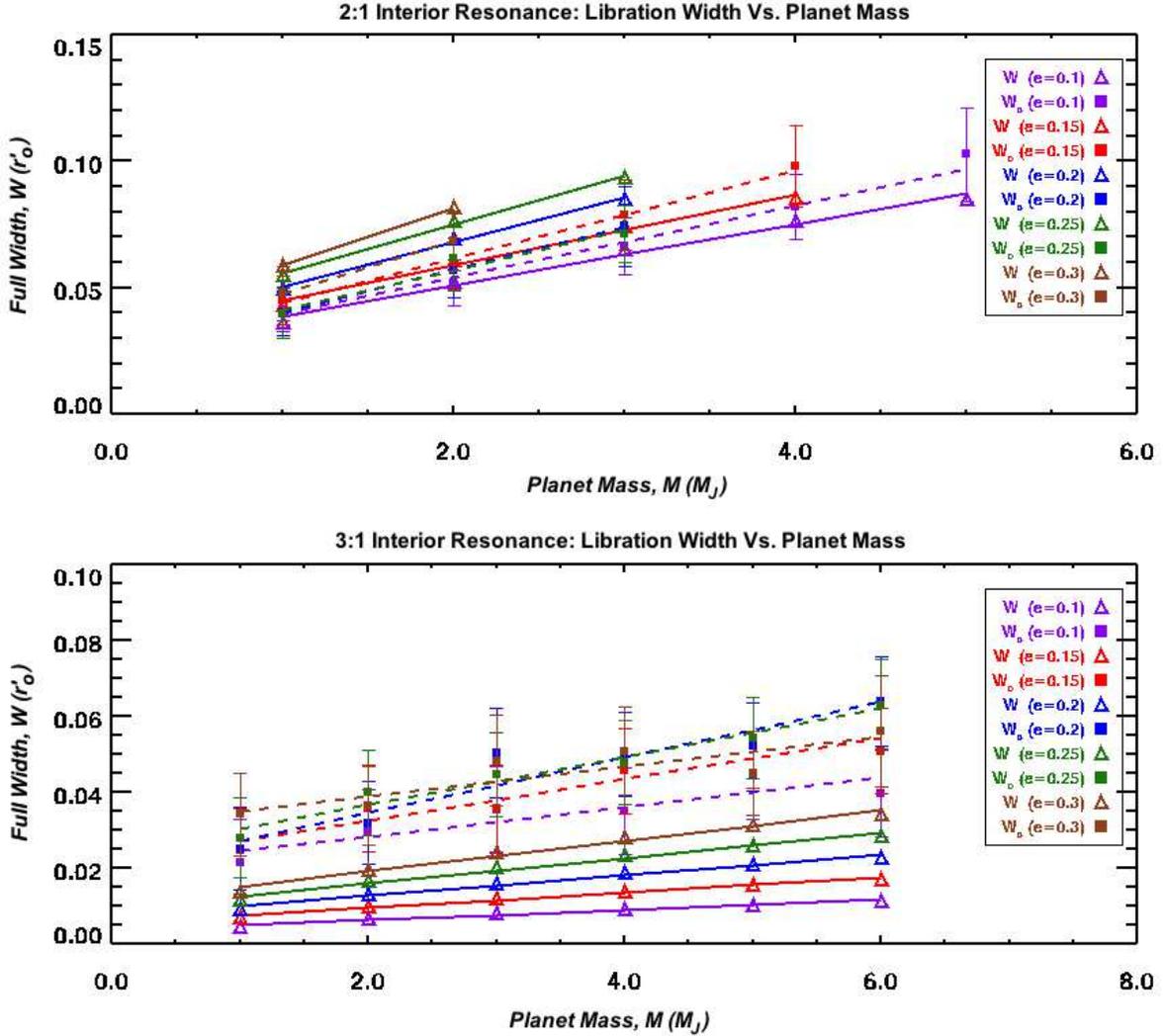}
    \caption{Change in MMR libration width with planet mass for 2:1 (top) and 3:1 (bottom) resonance with a planet exterior to the disk (i.e., interior resonance). The different colors represent different planet eccentricities used in the simulations. The solid and dashed lines are least-squares fits to the gap widths obtained analytically (triangles) and by Gaussian fitting to the gaps in particle distribution (squares), respectively.}
    \label{Fig:WvsM-IntRes}
\end{figure*}

\begin{figure*}
    \centering
    \includegraphics[totalheight=0.6\textheight]{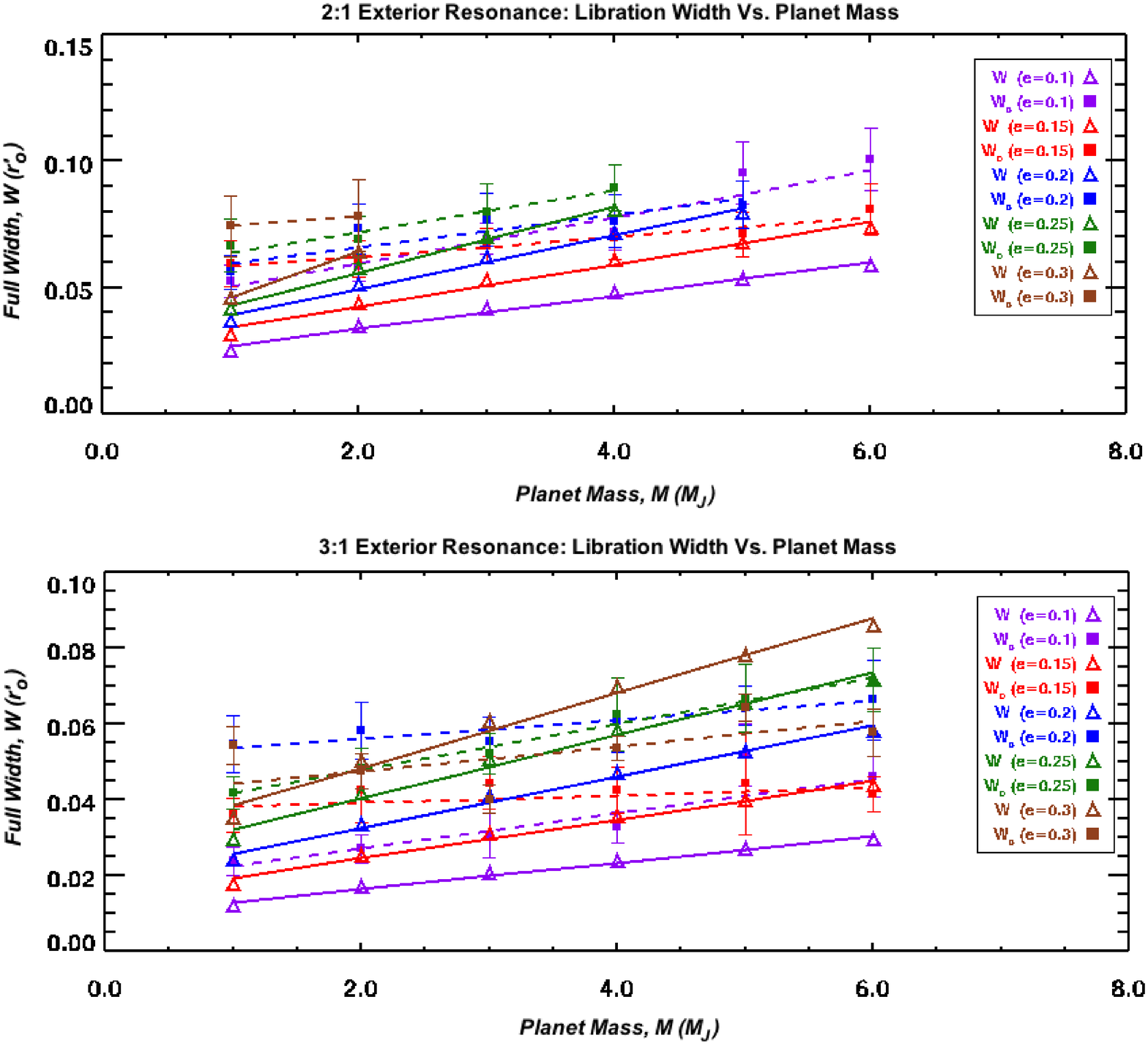}
    \caption{Same as Figure \ref{Fig:WvsM-IntRes} but for exterior resonance.}
    \label{Fig:WvsM-ExtRes}
\end{figure*}

More massive planets carve out wider resonance gaps in the disk. The change in gap width with planet mass is plotted in Figure \ref{Fig:WvsM-IntRes} for interior resonance and in Figure \ref{Fig:WvsM-ExtRes} for exterior resonance. The triangles show libration widths calculated analytically, while the squares are found by fitting Gaussians to the gaps as explained in Section \ref{Sec:Results}. The solid and dashed lines are our linear fits to the calculated values and those measured by the Gaussian fitting to the gaps, respectively; the different colors are for different planet eccentricities.

Our results show an increasing trend in gap width with increasing planet mass and eccentricity, as the theoretical calculation of the resonance width also suggests. Therefore, we propose a set of equations that allows the mass of the perturbing planet to be calculated for the different perturber eccentricities without the need to directly detect the planet or infer its mass through other means if (a) we can determine which MMR gap we observe in the disk and (b) a measure of the MMR gap width and eccentricity can be obtained observationally. The relation between an MMR gap width and the perturber's mass when a range of moderate planet eccentricity values is used is shown by the equations in Table \ref{Tab:WvsMP} for the 2:1 and 3:1 interior and exterior MMRs. These are drawn from least-squares fits to the values we obtain by Gaussian fitting to the simulation histograms where the gaps appear, as discussed in Section \ref{Sec:Results}. In these equations, $M$ is the planet mass (in $M_J$) and $W_o$ is the observed MMR gap width (in units of the distance between the star and the observed gap, ${r^\prime}_o$).

\begin{table*}
\centering
\caption{Equations Relating the Mass of a Perturber, $M$ in $M_J$, Having Various Orbital Eccentricities to the Observed Width of a Gap, $W_o$ in ${r^\prime}_o$ (the Observed Gap Location), at the 2:1 and 3:1 Interior and Exterior MMRs Drawn from Our Measurements of Gap Widths}
\label{my-label}
\begin{tabular}{|m{0.73cm}|m{3.5cm}|m{3.5cm}|m{3.5cm}|m{3.5cm}|@{}m{0pt}@{}}
\hline
\multicolumn{1}{|c|}{\multirow{2}{*}{\large{\textbf{e}}}} & 
\multicolumn{2}{c|}{\large{\textbf{Interior Resonance}}}  & 
\multicolumn{2}{c|}{\large{\textbf{Exterior Resonance}}} & \\ [10pt] \cline{2-5} 
\multicolumn{1}{|c|}{}            & 
\multicolumn{1}{c|}{\normalsize{\textbf{2:1}}}          & 
\multicolumn{1}{c|}{\normalsize{\textbf{3:1}}}          & 
\multicolumn{1}{c|}{\normalsize{\textbf{2:1}}}          & 
\multicolumn{1}{c|}{\normalsize{\textbf{3:1}}}          & \\ [10pt] \hline
\normalsize{\textbf{0.1}}                      & 
\normalsize{$M=\frac{1}{0.014}(W_o-0.025)$}    & 
\normalsize{$M=\frac{1}{0.004}(W_o-0.021)$}    & 
\normalsize{$M=\frac{1}{0.009}(W_o-0.041)$}    & 
\normalsize{$M=\frac{1}{0.005}(W_o-0.018)$}    & \\ [10pt] \hline
\normalsize{\textbf{0.15}}                     & 
\normalsize{$~~~~=\frac{1}{0.018}(W_o-0.026)$} & 
\normalsize{$~~~~=\frac{1}{0.005}(W_o-0.022)$} & 
\normalsize{$~~~~=\frac{1}{0.004}(W_o-0.054)$} & 
\normalsize{$~~~~=\frac{1}{0.001}(W_o-0.037)$} & \\ [10pt] \hline
\normalsize{\textbf{0.2}}                      & 
\normalsize{$~~~~=\frac{1}{0.017}(W_o-0.023)$} & 
\normalsize{$~~~~=\frac{1}{0.007}(W_o-0.020)$} & 
\normalsize{$~~~~=\frac{1}{0.006}(W_o-0.053)$} & 
\normalsize{$~~~~=\frac{1}{0.003}(W_o-0.051)$} & \\ [10pt] \hline
\normalsize{\textbf{0.25}}                     & 
\normalsize{$~~~~=\frac{1}{0.016}(W_o-0.025)$} & 
\normalsize{$~~~~=\frac{1}{0.006}(W_o-0.024)$} & 
\normalsize{$~~~~=\frac{1}{0.008}(W_o-0.056)$} & 
\normalsize{$~~~~=\frac{1}{0.006}(W_o-0.036)$} & \\ [10pt] \hline
\normalsize{\textbf{0.3}}                      & 
\normalsize{$~~~~=\frac{1}{0.021}(W_o-0.027)$} & 
\normalsize{$~~~~=\frac{1}{0.004}(W_o-0.031)$} & 
\normalsize{$~~~~=\frac{1}{0.004}(W_o-0.070)$} & 
\normalsize{$~~~~=\frac{1}{0.003}(W_o-0.041)$} & \\ [10pt] \hline
\end{tabular}
\label{Tab:WvsMP}
\end{table*}

Our measurements of gap widths for structures formed at the 2:1 interior MMR with a single planetary perturber are within 25\% of theoretical values. The difference is larger when the analysis is done on the 3:1 interior resonance gap, as there seems to be a systematic offset between the calculated and measured values for the gap width. We attribute the difference between the measured and calculated widths to the fact that the equations to calculate $\delta a^{\prime}_{max}$ presented in \cite{Murray99} are only first-order approximations when the eccentricities are greater than zero. Furthermore, in \textit{Paper I}, we saw spiral patterns forming in the disk when the planet was placed interior to the disk (i.e., exterior resonance) that we believed were due to Lindblad resonances generating from the 3:1 MMR. This makes defining the edges of the gaps more difficult in this case and may be the reason our results for the exterior resonance case shown in Figure \ref{Fig:WvsM-ExtRes} have an inconsistency in slope with the theoretical values, more so than in the interior resonance case. Nevertheless, we propose that the set of equations presented in this study (Table \ref{Tab:WvsMP}) can be used to estimate the mass of the planetary perturber to within 1 $M_J$.


\subsection{Disk Offset and Pericenter/Apocenter Glow}
\label{Sec:DiskOffset}

As the planet eccentricity increases, so does the forced eccentricity of the disk, causing a net offset in the overall particle distribution away from the central star. This offset is away from the direction of the forced pericenter of the disk particle orbits, confirming the findings by \cite{Wyatt99} discussed in Section \ref{Sec:PeriAlign_Offset} that a physical disk offset toward apocenter is to be expected if there is a perturbing body with nonzero orbital eccentricity. Therefore, we also find that the presence of a disk offset may be evidence of a planetary (or stellar) companion on an eccentric orbit.

Furthermore, we investigate the wavelength dependence of the pericenter/apocenter brightness variations, the "pericenter (or apocenter) glow." To do so, we bin particles in $x$ and $y$ and assign a flux to each bin, assuming that the particles emit as perfect blackbodies. The pixel values on opposing sides of the disk are then added and compared. We note that the pericenter-versus-apocenter glow depends on the wavelength of observation, as was found by \cite{Pan16}. The magnitude of the effect depends on the disk and star parameters. While a thorough study of this phenomenon is outside the scope of this paper, particularly since we have not included submicron dust in our simulations, we note that the pericenter/apocenter difference can easily reach several percent. For instance, a $1.0 ~M_J$ planet with $e=0.3$ placed 1 AU away from a debris disk orbiting a solar-mass star would result in $8 \%$ more flux from the apocenter side of the disk when observed at $1300 ~\mu m$. However, when the same disk is observed at $10 ~\mu m$, we find a pericenter glow of $10 \%$. Here we note again that resonant structures may not be visible in the observed disks if studied at submicron wavelengths due to the gaps being washed out by submicron-size dust as it migrates outward in the disk by stellar radiation pressure \citep{Kuchner10}.


\subsection{Simulating ALMA Observations}
\label{Sec:ALMA_Sim}

Whereas MMR gaps are clearly visible in our simulated disks, whether they can be detected in a telescopic image of a debris disk depends largely on current observing capabilities. The technology is reaching the point at which we should start seeing a variety of structures, including the resonance gaps discussed in this paper, as ALMA images of second-generation disks emerge. Therefore, we discuss the observability of MMR gaps as seen by powerful interferometers such as ALMA. For this purpose, we use the Common Astronomy Software Applications (CASA) offered by the National Radio Astronomy Observatory (NRAO) to simulate ALMA observations  \citep{CASASim}.

We use as our fiducial example the \textit{AU Microscopii} debris disk, which has already been well studied with ALMA \citep{MacGregor13}. Synthetic images of our simulated disks are created on the assumption that they are the same size (140 AU radius), distance (9.91 pc), and overall brightness (7.14 mJy) as the one around AU Mic. AU Mic is an $\sim$ 10 Myr old M-type star with $R=0.83 ~R_\odot$ and $T=3600 ~K$ \citep{Matthews15} that has an edge-on debris disk first discovered by \cite{Kalas04}. 

Our simulated disk is taken to be optically thin and composed of perfect blackbodies emitting at the local equilibrium temperature. We then use CASA to determine how our simulated disks would appear if observed with the same resolution used to image the AU Mic disk with ALMA at 230 GHz or 1.3 mm \citep[see][]{MacGregor13} if they were to be viewed face-on. This would correspond to a resolution of 0.\arcsec6 or about 6 AU. 

\begin{figure*}
    \centering
    \subfigure[]{%
    \includegraphics[totalheight=0.43\textheight]{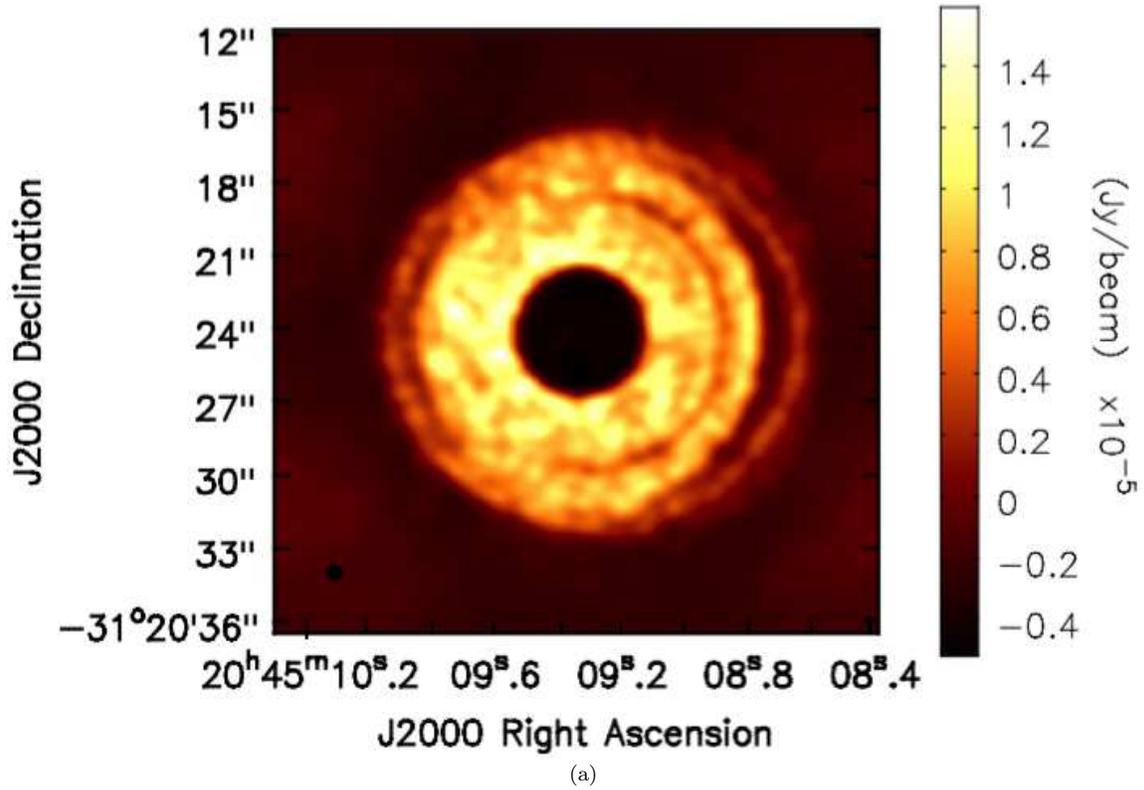}
    \label{Fig:ALMAOuter}}
\hfill
    \subfigure[]{%
    \includegraphics[totalheight=0.43\textheight]{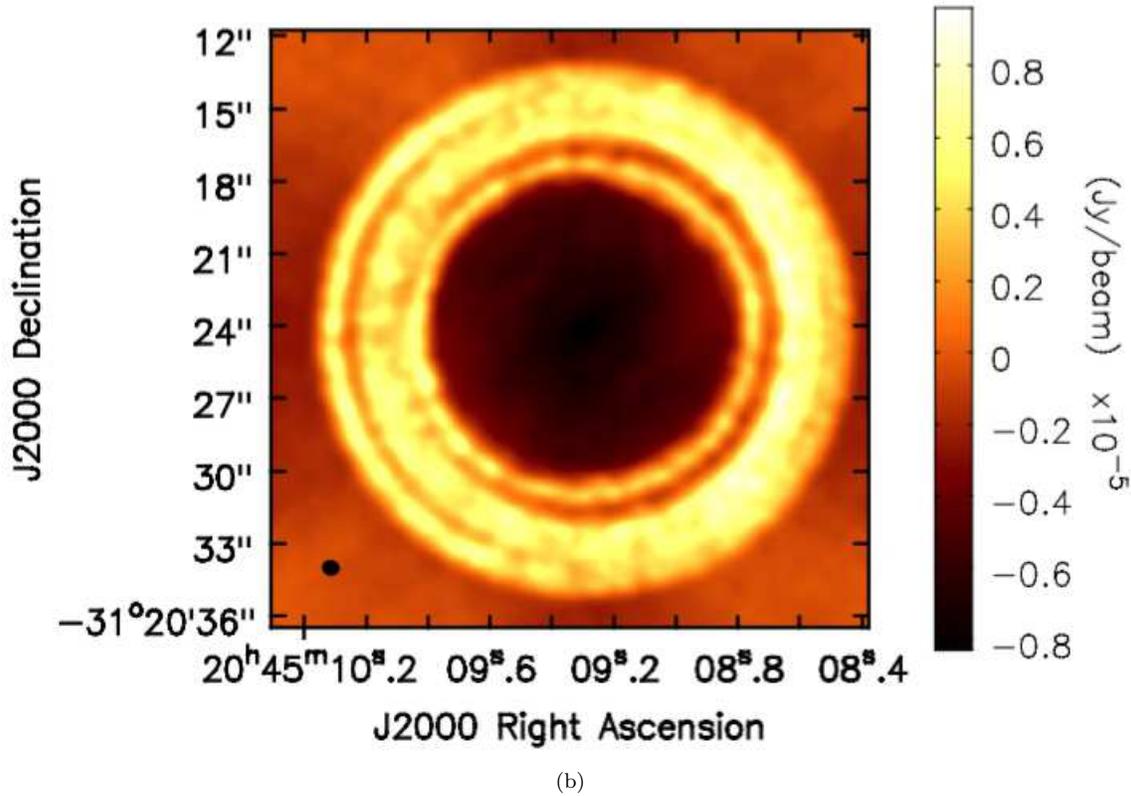}
    \label{Fig:ALMAInner}}
\caption{Using the CASA simulator, this is how the disks in Figures \ref{Fig:outerE0.1m3-disk} (top) and \ref{Fig:innerE0.1m3-disk} (bottom) would look after beam deconvolution if they were placed at the AU Mic distance and observed with the same resolution used in observing its debris disk (0.\arcsec6). The arcs seen in these images correspond to gaps formed at the 2:1 and 3:1 interior (top) and exterior (bottom) MMRs with a planet. Although the 2:1 gap has a better contrast compared to the 3:1 gap in the simulated images, both gaps will be visible. The color bar shows the flux in (Jy beam\!\! $^{-1}$) $\times 10^{-5}$. The synthesized beam is shown by a black ellipse in the lower left corner and is $0.\arcsec68 \times 0.\arcsec60$.}
\label{Fig:ALMASims}
\end{figure*}

When ALMA was used for the first time to observe the debris disk around AU Mic in 2012, there were only 20 operational 12 m antennas. However, we utilize all 50 antennas available in the 12 m array to make our simulated images to achieve the desired resolution. Furthermore, we set the integration time to 10 s per pointing and assume that the disk is observed for a total of 4 hr. The R.A. and decl. of the source are also taken from \cite{MacGregor13}: $\alpha=20^h45^m09^s.34$ and $\delta=-31^{\degr}20^\prime24\arcsec.09$ (J2000). We take the column density of the precipitable water vapor to be 1.796 mm, which is to be expected for more than half the observations at the ALMA site, and use the recommended values for the sky temperature, opacity, and system temperature of 22.558 K, 0.092, and 103.542 K, respectively. We also use dual polarization and a 7.5 GHz bandwidth, recommended for continuum observations with ALMA. Figure \ref{Fig:ALMASims} illustrates an example of two beam-deconvolved images that we made with CASA; the top and bottom figures show the same disks as in Figures \ref{Fig:outerE0.1m3-disk} and \ref{Fig:innerE0.1m3-disk}, respectively. In both examples, the MMR structures in the simulated disks are easily visible.

\begin{figure*}
    \centering
    \includegraphics[totalheight=0.4\textheight]{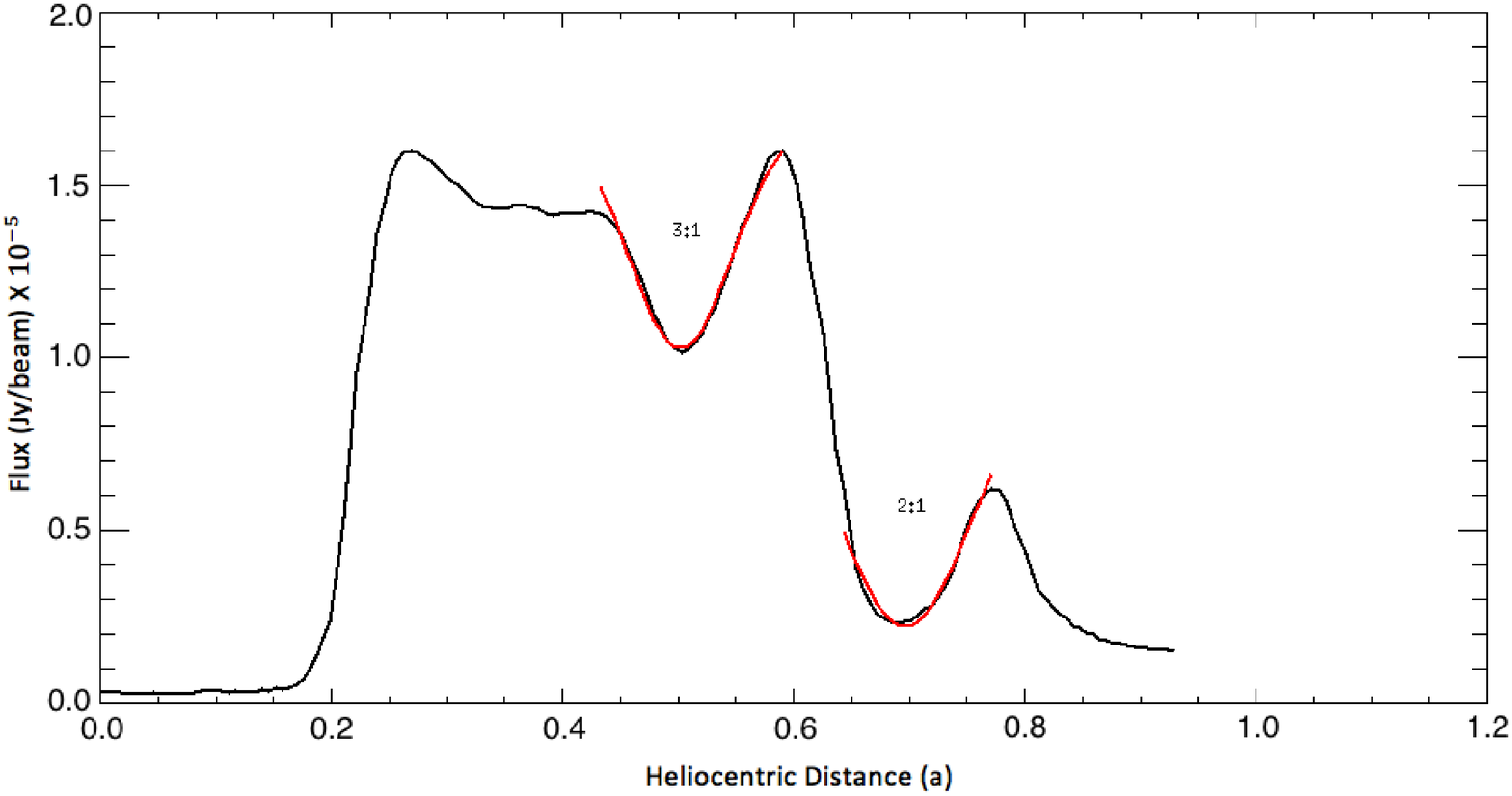}
    \caption{Surface brightness profile of the disk shown in Figure \ref{Fig:ALMAOuter} along the major axis. A Gaussian function is used to fit each gap to measure its depth. The gap edge-to-center contrast is $60\%$ for the 2:1 gap and $30\%$ for the 3:1, indicating that both gaps are deep enough to be detectable.}
    \label{Fig:Contrast}
\end{figure*}

In order to asses the observability of the structures in our simulated disks, we calculated the edge-to-center contrast for each gap and noted that for the 2:1 and 3:1 gaps in our simulated images, the contrast is about $60\%$ and $30\%$, respectively (see Figure \ref{Fig:Contrast}). This means that both the 2:1 and 3:1 gaps in Figures \ref{Fig:ALMAOuter} and \ref{Fig:ALMAInner} would produce high contrast and should be visible by ALMA. Therefore, we argue that given the high sensitivity and resolving power that can be achieved with ALMA, the structures discussed in this paper are, in fact, within current detectability limits.


Although debris disks have shown a wide range of structures, current images of resolved debris disks at long wavelengths generally have low signal-to-noise ratios and do not reveal much detail about these structures. We would need higher-resolution images to be able to observe and measure the structures that we discussed in this work. However, recent images of protoplanetary and transitional disks, especially those obtained by ALMA—such as the images of the disks around HL Tau, TW Hydrae, and HD 141569—are very promising for the future of detecting detailed structures such as those formed by resonances. In fact, some gaps in the HL Tau disk may be due to MMRs with some of its potentially embedded planets, specifically the gaps at roughly 38 and 46 AU, which could be due to 3:2 and 2:1 MMRs with the planet at 29 AU (Tabeshian \& Wiegert 2017b, in preparation). However, to our knowledge, there have thus far been no observations of debris disks with structures that resemble what we have seen theoretically for MMR gaps.


\section{Summary and Conclusions}
\label{Sec:SummConc}

We extended our study of gaps formed through resonant interactions of a single planet with a gas-poor dynamically cold debris disk, presented in an earlier paper \citep{Tabeshian16}, to include systems in which the planet has moderate orbital eccentricity. Gravitational perturbation of the particles by a planet forms gaps whose locations correspond to the MMRs with the planet.

Unlike gaps cleared by planets around their orbits, we found that the MMR gaps, formed away from the orbits of the planets, are not azimuthally symmetric about the star. For the 2:1 MMR, a planet orbiting exterior to the disk leaves its resonance imprint as two arc-shaped gaps at inferior and superior conjunctions but forms a single arc at opposition if placed interior to the disk. This difference allows observers to distinguish between interior and exterior resonances solely based on the shape of the 2:1 gap.

We thus provided a simple procedure for determining the mass, semimajor axis, and eccentricity of the planetary perturber from single-epoch measurements of a debris disk. If multi-epoch observations are available, the determination becomes easier. Nevertheless, the planetary parameters can be determined from the resonant structures even if the planet itself remains unseen by analyzing the resonance gaps as follows.

(A) The eccentricity at the center of an MMR gap can be measured by least-squares fitting of ellipses to the gap edges (Section \ref{Sec:e_pl}). 

(B) The distance between a gap and the host star can be determined observationally if the distance to the system being studied is known, which is often the case for nearby debris disks that have been observed. This information, together with the eccentricity of the gap and the true anomaly of its center, can help calculate the gap's semimajor axis, $a^\prime$, using Equation \ref{Eq:Ellipse}.

(C) If we can determine which resonance gap is observed in the disk, calculating the planet's semimajor axis is trivial and can be done using Equation \ref{Eq:MMR_Def_a}. Alternatively, the planet's semimajor axis can be found if its orbital motion is detected in multi-epoch observations of the disk (Section \ref{Sec:a_pl}).

(D) Once the semimajor axes of the planet and gap are found, the eccentricity of the planet can be determined using the forced eccentricity at the center of the gap and Equation \ref{Eq:Forced_e}. This is true since, in a dynamically cold debris disk where disk particles can be assumed to have zero or negligible free eccentricities, orbital eccentricity anywhere in the disk is defined by the forced eccentricity induced by the planet at that location.

(E) Finally, since the libration width of an MMR gap is related to the perturber's mass and eccentricity, a measurement of the gap width can help determine the mass of the planet using the formulae that we presented in this work (see Table \ref{Tab:WvsMP}).

In addition to the 2:1 gap, we found that increasing the perturber's eccentricity resulted in formation of a second gap at the 3:1 MMR that forms a single arc. Increasing the perturber's orbital eccentricity also resulted in formation of higher-order resonance gaps in the disk. Furthermore, we noted that while the 2:1 gap orbits the star at the same rate as the planet, the 3:1 gap remains stationary in the inertial frame. It appears at apocenter for interior and at pericenter for exterior MMRs. This difference can be important if multi-epoch observations of the disk are available.

Furthermore, we independently confirmed the result of \cite{Pan16} for the wavelength dependence of the apocenter/pericenter glow phenomenon, which is a trade-off between a larger number of particles at apocenter and enhanced flux caused by the disk offset away from pericenter in debris disks that are perturbed by a planet with nonzero orbital eccentricity.

By means of the CASA simulator, we showed that resonance structures should be detectable in images of suitable debris disks using ALMA or other high-resolution facilities. We conclude that the analysis of MMR gaps in extrasolar debris disks is a useful indirect technique to not only detect but also characterize extrasolar planets.

\acknowledgments

The authors wish to thank Brenda Matthews for assistance with CASA. We also thank the anonymous referee for the valuable comments and suggestions that we received. This work was supported in part by the Natural Sciences and Engineering Research Council of Canada (NSERC).


\bibliography{main} 

\end{document}